\documentclass[10pt,a4paper]{article}
\usepackage{amsmath,amssymb,graphicx,microtype}

\def\be{\begin{eqnarray}}
\def\ee{\end{eqnarray}}
\def\nn{\nonumber}

\def\tr{{\rm tr}\,}

\def\l[{\phantom.[}

\newcommand{\beq}{\begin{equation}}
\newcommand{\eeq}{\end{equation}}
\newcommand{\bal}{\begin{align}}
\newcommand{\eal}{\end{align}}

\newcommand{\ket}[1]{\left| #1 \right\rangle }
\newcommand{\braket}[2]{\left\langle #1 | #2 \right\rangle }
\newcommand{\vl}{\vec{\lambda}}
\newcommand{\vm}{\vec{\mu}}
\newcommand{\vn}{\vec{\nu}}

\newcommand{\lo}{\lambda^{(1)}}
\newcommand{\lt}{\lambda^{(2)}}
\newcommand{\mo}{\mu^{(1)}}
\newcommand{\mt}{\mu^{(2)}}
\newcommand{\Xo}{X^{(1)}}
\newcommand{\Xt}{X^{(2)}}

\hoffset= - 1.0in         

\begin{document}

\title{\vspace{.1cm}{\Large {\bf Anomaly in $\mathcal{RTT}$ relation
      for DIM algebra\\
      and network matrix models
}\vspace{.2cm}}
\author{
{\bf Hidetoshi Awata$^a$}\footnote{awata@math.nagoya-u.ac.jp},
\ {\bf Hiroaki Kanno$^{a,b}$}\footnote{kanno@math.nagoya-u.ac.jp},
\ {\bf Andrei Mironov$^{c,d,e,f}$}\footnote{mironov@lpi.ru; mironov@itep.ru},
\ {\bf Alexei Morozov$^{d,e,f}$}\thanks{morozov@itep.ru},\\
\ {\bf Andrey Morozov$^{d,e,f,g}$}\footnote{andrey.morozov@itep.ru},
\ {\bf Yusuke Ohkubo$^a$}\footnote{m12010t@math.nagoya-u.ac.jp}
\ \ and \ {\bf Yegor Zenkevich$^{d,f,h,i,j,k}$}\thanks{yegor.zenkevich@gmail.com}}
\date{ }
}

\maketitle

\vspace{-6.5cm}

\begin{center}
\hfill FIAN/TD-24/16\\
\hfill IITP/TH-18/16\\
\hfill ITEP/TH-26/16\\
\hfill INR-TH-2016-041
\end{center}

\vspace{4.3cm}

\begin{center}
$^a$ {\small {\it Graduate School of Mathematics, Nagoya University,
Nagoya, 464-8602, Japan}}\\
$^b$ {\small {\it KMI, Nagoya University,
Nagoya, 464-8602, Japan}}\\
$^c$ {\small {\it Lebedev Physics Institute, Moscow 119991, Russia}}\\
$^d$ {\small {\it ITEP, Moscow 117218, Russia}}\\
$^e$ {\small {\it Institute for Information Transmission Problems, Moscow 127994, Russia}}\\
$^f$ {\small {\it National Research Nuclear University MEPhI, Moscow 115409, Russia }}\\
$^g$ {\small {\it Laboratory of Quantum Topology, Chelyabinsk State University, Chelyabinsk 454001, Russia }}\\
$^h$ {\small {\it Institute of Nuclear Research, Moscow 117312, Russia
  }}\\
$^i$ {\small {\it Physics Department, Moscow State University, Moscow 117312, Russia
}}\\
$^j$ {\small {\it Dipartimento di Fisica, Universit\`a di Milano-Bicocca,
Piazza della Scienza 3, I-20126 Milano, Italy}}\\
$^k$ {\small {\it INFN, sezione di Milano-Bicocca,
I-20126 Milano, Italy
  }}
\end{center}

\vspace{.5cm}

\begin{abstract}
  We discuss the recent proposal of arXiv:1608.05351 about
  generalization of the $RTT$ relation to network matrix models.  We
  show that the $RTT$ relation in these models is modified by a
  nontrivial, but essentially abelian anomaly cocycle, which we explicitly
  evaluate for the free field representations of the quantum toroidal
  algebra.  This cocycle is responsible for the braiding, which
  permutes the external legs in the $q$-deformed conformal block and
  its $5d/6d$ gauge theory counterpart, i.e. the non-perturbative
  Nekrasov functions. Thus, it defines their modular properties and
  symmetry. We show how to cancel the anomaly using a construction somewhat
  similar to the anomaly matching condition in gauge theory. We also describe
  the singular limit to the affine Yangian ($4d$ Nekrasov functions),
  which breaks the spectral duality.
\end{abstract}


\paragraph{1. The Ding-Iohara-Miki algebra (DIM),} a quantum
deformation of the toroidal algebra with two central charges
\cite{DI}-\cite{Awata:2016mxc} is known to be the underlying symmetry
of network matrix models \cite{Morozov:2015xya}-\cite{AKTMMMOZ}, which
describes the Seiberg-Witten-Nekrasov theory \cite{SW,GKMMM,Nek} at
the maximally general topological string level
\cite{top}-\cite{Awata:2008ed}. It also plays the central role in the
AGT correspondence \cite{AGT}-\cite{AGT5d}. In this work we will focus
on the simplest free field representations of the DIM algebra for
which a detailed description is known~\cite{F}.  Technically, the main
objects in this approach are the triple vertices: the {\bf
  intertwiners} of the DIM algebra~\cite{AFS}:
\begin{align}
  \label{eq:1}
  \Psi^{\lambda}(N,u|z)
  &=  \parbox{3cm}{\includegraphics[width=3cm]{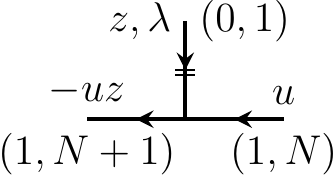}}
  \sim\ : e^{\Phi(z)}\prod_{\Box\in \lambda} e^{\phi(z_\Box)-\phi(z_\Box/t)} :
  \notag\\
  \notag\\
    \Psi^*_\lambda(N,-uw|w)
  &= \parbox{3cm}{\includegraphics[width=3cm]{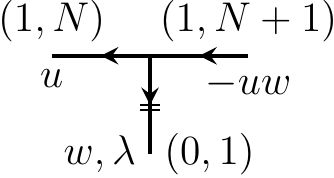}}
  \sim\ : e^{-\Phi^*(w)}\prod_{\Box\in \lambda} e^{-\phi^*(w_\Box)+\phi^*(w_\Box/t)}:
\end{align}
which are made from the free fields
\begin{equation}
\phi(z) = \sum_{n\geq 1} \frac{1}{n} \Big(a_{-n} z^n - a_n
z^{-n}\Big), \qquad \qquad
\phi^*(z) = \sum_{n\geq 1} \frac{1}{n} \Big(a_{-n}^* z^n - a_n^* z^{-n}\Big)
\end{equation}
and act as operators on the Fock space in the ``horizontal''
direction. Here $z_\Box$ means that the spectral parameter $z$ is
scaled with powers of $q$ and $t$ in a way depending on the box of the
Young diagram $\lambda$, see Appendix A.  The difference between
annihilation/creation operators $a_{\pm n}$ and $a^*_{\pm n}$
disappears in the $4d$ limit $q,t\to 1$, which, however, is a little
tricky, see Appendices A and D at the end of this paper.  Most
importantly, $\Psi$ and $\Psi^*$ contain ordinary vertex operators
together with all the necessary screening charges.

Each leg in the picture corresponds to a Fock representation
$\mathcal{F}_u^{(k,l)}$ of the DIM algebra.  The representations are
labeled by the complex spectral parameters and the integer-valued
slope vectors, which determine the values of two DIM central charges.
Both the slope vectors and the spectral parameters satisfy the obvious
balancing conditions at every vertex. The intertwiner $\Psi(N,u|z)$
(resp.\ $\Psi^{*}(N,-uz|z)$) maps from the tensor product of two Fock
representations $\mathcal{F}_u^{(1,N)}\otimes \mathcal{F}_z^{(0,1)}$
into a single Fock representation $\mathcal{F}_{-uz}^{(1,N+1)}$
(resp.\ vice versa).  We call Fock representations with slopes of the
form $(1,N)$ ``horizontal'' and those with the slope $(0,1)$
``vertical'' and draw them accordingly.  To make the presentation
simpler, we do not distinguish between the horizontal representations with
different slopes in our pictures (as shown e.g.\ in Eq.~\eqref{eq:3}).
We also omit the slope argument $N$ in the intertwiners when it is
clear on which representation the operator acts.

\paragraph{2. Balanced network matrix model} is a correlator (matrix
element in the product of Fock spaces) of the $T$-operators --- the
bilinear combinations of $\Psi$-intertwiners --- which geometrically
correspond to resolved conifolds:
\begin{equation}
\label{eq:3}
\mathcal{T}^{\alpha}_{\mu}(u| z, w) =  \Psi^{*}_{\mu}(0,-uz|w)
\Psi^{\alpha}(0,u|z) =
\parbox{8cm}{
  \includegraphics[width=8cm]{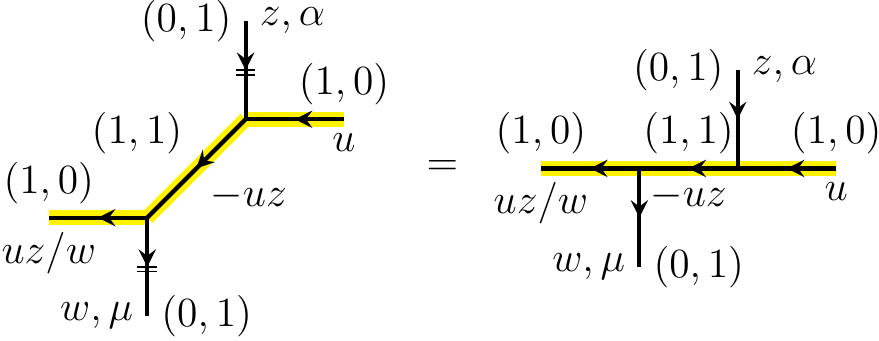}
  }
\end{equation}
As already mentioned, we use the second picture where all non-vertical
slopes are drawn horizontally to emphasize that the free field (the
Fock space, drawn in blue in Eq.~\eqref{eq:3}) remains the same along
the entire horizontal line.

Composing the $T$-operators, one can build vertical
``strip'' geometries $\mathcal{V}$. The simplest strip contains two
$T$-operators and corresponds to the screened vertex
operator of the $q$-deformed Virasoro algebra:
\begin{equation}
  \mathcal{V}^{\alpha}_{\beta}(u,v|z_1, z_2 , z_3) = \parbox{4.5cm}{\includegraphics[width=4.5cm]{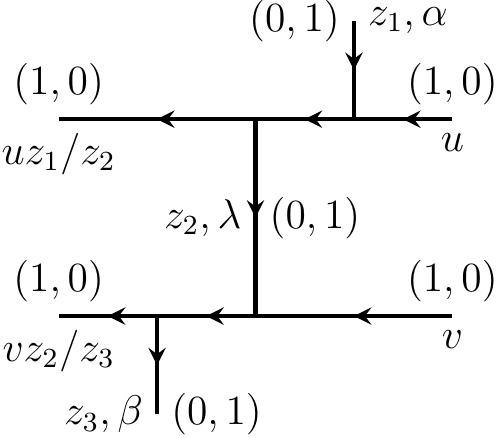}} =
  \sum_\lambda
  \begin{array}{c}
    \mathcal{T}^{\alpha}_{\lambda}(u|z_1, z_2)\\
    \otimes\\
  \mathcal{T}^{\lambda}_{\beta}(v|z_2, z_3)
  \end{array}
 =
  \sum_\lambda  \begin{array}{c}
    \Psi^*_\lambda(-uz_1 |z_2)\Psi^{\alpha}(u|z_1) \\
    \otimes \\
    \Psi^*_\beta(-v z_2|z_3)\Psi^{\lambda}(v|z_2)
  \end{array} \label{eq:15}
\end{equation}
Longer combinations contain $m$ $T$-operators and reproduce the
screened vertex operators of the $q$-deformed $W_m$-algebra:
\begin{equation}
  \mathcal{V}^{(m)}{}^{\alpha}_{\beta} (u_1,\ldots, u_m|z_1,
  z_2,\ldots, z_{m+1})= \parbox{5cm}{\includegraphics[width=5cm]{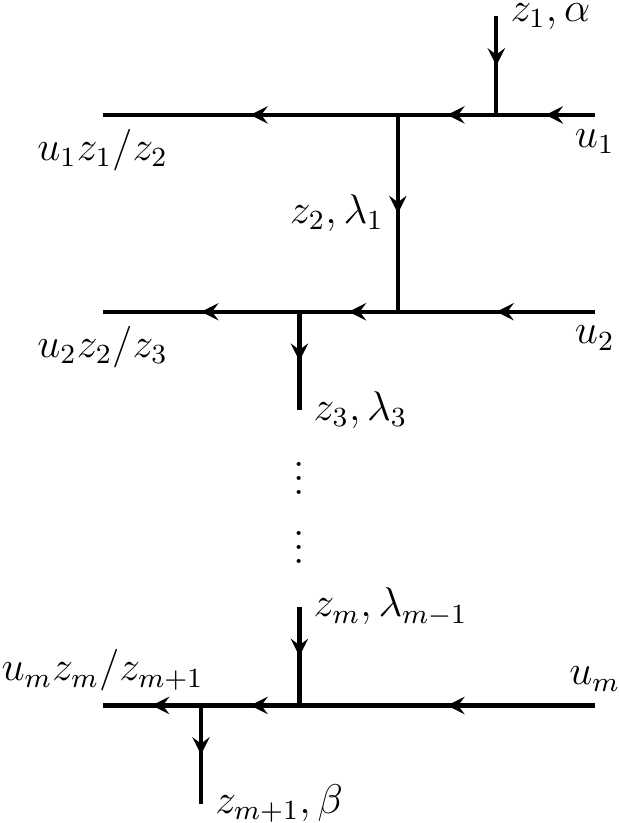}} =
  \sum_{\lambda_1,\ldots,\lambda_{m-1}}   \begin{array}{c}
    \mathcal{T}^{\alpha}_{\lambda_1}(u_1|z_1,z_2)\\
    \otimes\\
    \mathcal{T}^{\lambda_1}_{\lambda_2}(u_2|z_2, z_3)\\
    \otimes\\
    \vdots\\
    \otimes\\
    \mathcal{T}^{\lambda_{m-1}}_{\beta}(u_m|z_m, z_{m+1})
  \end{array}
  \label{2mfold}
\end{equation}
$\mathcal{V}^{(m)}{}^{\alpha}_{\beta}$ acts on the products of $m$
Fock spaces.  The operators of the $W_m$-algebra are generated from
the $(m-1)$-coproduct of a single rasing operator $x^{+}(z)$ of DIM~\cite{FHSSY-kernel}.

\paragraph{3. $R$-matrices.}
As in any quantum group \cite{DJ,FRT,DrinR}, in the DIM case, there
exists the group element $g$ satisfying the $RTT$ relation \cite{Fad}
\begin{equation}
  (I\otimes g)(g\otimes I) ={\cal R} ( g \otimes I) (I\otimes g){\cal R}^{-1}
\label{Rgg}
\end{equation}
where the universal $R$-matrix ${\cal R}$ satisfies the Yang-Baxter
relations
\begin{equation}
  \mathcal{R}^{(12)} \mathcal{R}^{(13)} \mathcal{R}^{(23)} = \mathcal{R}^{(23)} \mathcal{R}^{(13)} \mathcal{R}^{(12)}
\end{equation}
The universal $R$-matrix is triangular in the sense that it is an
element of the product of the universal enveloping algebras
$U_q(\mathfrak{b}_{+}) \otimes U_q(\mathfrak{b}_{-})$, where
$\mathfrak{b}_{\pm}$ are positive and negative Borel subalgebras.

When one defines the $R$-matrix, there are two sources of confusion,
which we discuss in turn:
\begin{enumerate}
\item[\bf 1)] The choice of the positive and negative Borel
  subalgebras $\mathfrak{b}_{\pm}$ is irrelevant for most ordinary
  situations, since different choices are conjugate to each
  other. However, in the DIM algebra case this equivalence requires a
  more careful treatment.

  To see this, let us start with a simplifier example of quantum
  affine algebra \cite{DJ} $U_q(\widehat{\mathfrak{sl}}_2)$. The roots
  of the algebra are $E_m$, $F_m$ and $H_m$ for $m \in \mathbb{Z}$
  ($H_0$ is not included). There are two natural choices of the Borel
  subalgebras. The first one is $\mathfrak{b}_{+} = \langle E_m , m
  \geq 0, H_n, n \geq 1, F_k, k \geq 1\rangle$. This leads to the
  standard universal $R$-matrix \cite{DrinR}, which in the evaluation
  representation \cite{eval} associated with the fundamental
  representation of the finite-dimensional algebra $\mathfrak{sl}_2$
  looks like \cite{Raz}
  \begin{equation}
    \label{eq:31}
    \mathcal{R}(x) = \left(
      \begin{array}{cccc}
        1 & 0 & 0 & 0\\
        0 & \frac{q (1 - x)}{1 - q^2 x} & \frac{(1 - q^2)x}{1 - q^2
          x}  & 0\\
        0 & \frac{(1 - q^2)}{1 - q^2
          x}  & \frac{q (1 - x)}{1 - q^2 x} & 0\\
        0 & 0 & 0 & 1
      \end{array}
    \right)
  \end{equation}
  This is not triangular in the sense of conventional
  $\mathfrak{sl}_2$ (as a finite matrix), because the positive roots
  of the affine algebra contain the modes of both positive and
  negative roots of $\mathfrak{sl}_2$. There is also a different
  choice of the Borel subalgebras $\mathfrak{b}^{\perp}_{\pm}$, in the
  spirit of Drinfeld's ``new realization'' \cite{Drinnew} of
  $U_q(\widehat{\mathfrak{sl}}_2)$. If we set
  $\mathfrak{b}_{+}^{\perp} = \langle H_m, m \geq 1, E_n, n \in
  \mathbb{Z} \rangle$, then the resulting $R$-matrix is
  upper-triangular in the conventional sense. The two $R$-matrices are
  conjugate to each other, with the conjugation matrix being given by
  $\prod_{\alpha} "\exp"\Big(E_\alpha\otimes F_\alpha\Big)$, where the
  product is taken over the roots $E_n$ with $n < 0$. For more details
  see Appendix E.

  In the DIM algebra case, the two choices of the Borel subalgebra,
  ``vertical'' and ``horizontal'' are related by the \emph{spectral
    duality} automorphism $\mathcal{S}$ \cite{Miki,specdu} (a proof is
  given in Lemma A.5 of~\cite{FJMM} and below we give explicit
  examples).

  The $T$-operator in this construction should be considered
  as the universal $R$-matrix taken in the tensor product of the
  horizontal \emph{and} vertical representations. The operators thus
  constructed automatically satisfy the $RTT$ relations. However, the
  $T$-operator, which we consider is obtained from an altogether
  different considerations: it is the topological string amplitude on
  resolved conifold, the basic building block of the network matrix
  model, and is given by the combination of two DIM
  intertwiners. There is \emph{a priori} no guarantee that this
  $T$-operator will satisfy the $RTT$ relations. Indeed the relations
  become \emph{anomalous.}  However, as we will see in the next
  section, the anomaly is very ``mild'': it is just a scalar
  multiplier depending on the spectral parameters.

  Most importantly, the choice of $R$-matrix is connected with the
  choice of the type of the Fock representations on which the matrix
  acts.  The ``vertical'' $R$-matrix acts on the vertical Fock
  representations, while its action on the horizontal Fock spaces is
  undefined, since the $R$-matrix contains an infinite number of both
  positive and negative (from the horizontal point of view) roots and
  there is no way to make the action finite. Of course, for the
  ``horizontal'' $R$-matrix the situation is reversed and it can
  safely act on the horizontal Fock modules, while the vertical
  representations are off-limits to it.

  The ``vertical'' $R$-matrix $\mathcal{R}$ was studied in~\cite{FJMM}
  (see also \cite{FJMM-Bethe,Rmat}), and the ``horizontal'' one
  $\hat{\mathcal{R}}$ was written out in~\cite{Awata:2016mxc} using
  the generalized Macdonald polynomials (note that we used there the
  $RTT$-relation normalized to the contribution of the empty Young
  diagrams, or of the highest weight vectors, thus the anomaly was not
  visible). In this paper, we use both of these $R$-matrices (though
  in different situations, see pictures in the next sections).

\item[\bf 2)] As any universal $R$-matrix, the DIM $R$-matrix tends to
  identity as the quantization parameter tends to identity. There is
  also a different convention on numbering the tensor indices in the
  $R$-matrix. One can multiply the original universal $R$-matrix by
  the permutation operator acting on the tensor indices of
  representations (but not on the spectral parameters):
  \begin{equation}
    R = P {\cal R}
  \end{equation}
  The new $R$-matrix tends to the permutation in the classical limit
  and satisfies the Reidemeister (Hecke- or braid-algebra)-like
  relations
  \begin{equation}
    \label{eq:6}
    R^{(12)} \left( \frac{z_1}{z_2} \right) R^{(23)}\left(
      \frac{z_1}{z_3} \right) R^{(12)} \left( \frac{z_2}{z_3} \right) =
    R^{(23)}\left( \frac{z_2}{z_3} \right) R^{(12)}\left(
      \frac{z_1}{z_3} \right) R^{(23)} \left( \frac{z_1}{z_2} \right)
  \end{equation}
  Notice that the indices of the spectral parameters do not match with
  the tensor indices. One can also introduce \emph{two} more
  $R$-matrices, where the permutation acts on the spectral parameters
  and possibly on the tensor indices.

  The original $R$-matrix computed in~\cite{Awata:2016mxc} was of the
  universal type (i.e. tended to identity as $q\to 1$). This is
  related to the choice of basis in the tensor product of Fock
  representations. Correspondingly, there are two ways to write the
  $RTT$ relations. In the next section, we will write the $RTT$
  relations assuming that the $R$-matrix is of the ``braid'' type. In
  the language of pictures, the $R$-matrix exchanges the two parallel
  legs. Then, it depends on whether we order all indices according to
  the ordering of the \emph{legs} on the diagram, or according to the
  ordering of the \emph{places} on which the legs reside, i.e.\ when
  the two legs are exchanged, the corresponding indices are either
  exchanged or stay at their places (see e.g.\ Eq.~\eqref{Rvert}).
\end{enumerate}

\paragraph{4. The claim of \cite{Awata:2016mxc}}
was that the role of the universal group element $g$ in the Fock
representations is played by the $\Psi$-bilinear combination
$\mathcal{T}^{\alpha}_{\mu}$, moreover, in this case, as a
manifestation of the spectral duality, there are two spectral dual
$R$-matrices. The first one acts in the vertical channel and is a
matrix explicitly depending on four Young-diagrams
\begin{equation}
  \label{eq:30}
  \parbox{14cm}{\includegraphics[width=14cm]{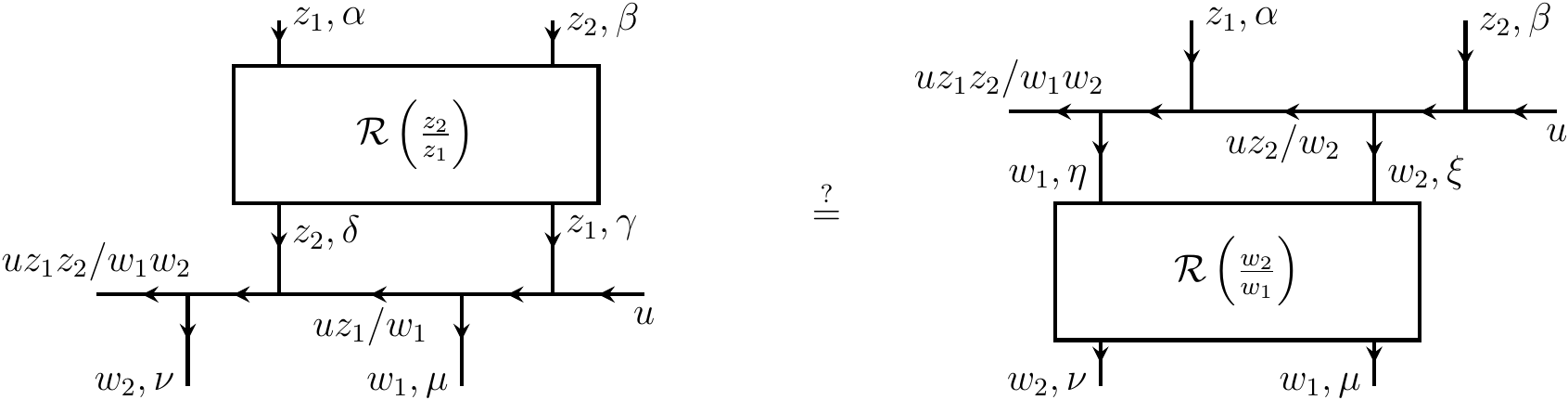}}
\end{equation}
\begin{equation}
  \sum_{\gamma,\delta} {\cal R}^{\alpha\beta}_{\delta\gamma}\left(
    \frac{z_2}{z_1} \right)
  \mathcal{T}^{\delta}_{\nu}\left(\frac{uz_1}{w_1}\Big|z_2, w_2\right)
  \mathcal{T}^{\gamma}_{\mu}(u|z_1,w_1) \stackrel{?}{=}
  \sum_{\xi,\eta}
  \mathcal{T}^{\alpha}_{\eta}\left(\frac{uz_2}{w_2}\Big|z_1,
    w_1\right) \mathcal{T}^{\beta}_{\xi}(u|z_2, w_2) {\cal
    R}^{\eta\xi}_{\nu\mu} \left( \frac{w_2}{w_1} \right)
\label{Rvert}
\end{equation}
The second $R$-matrix, which we denote through
$\hat{\mathcal{R}}$ acts in the horizontal channel and is an operator
in the tensor product of two Fock spaces:
\begin{equation}
  \parbox{12cm}{\includegraphics[width=12cm]{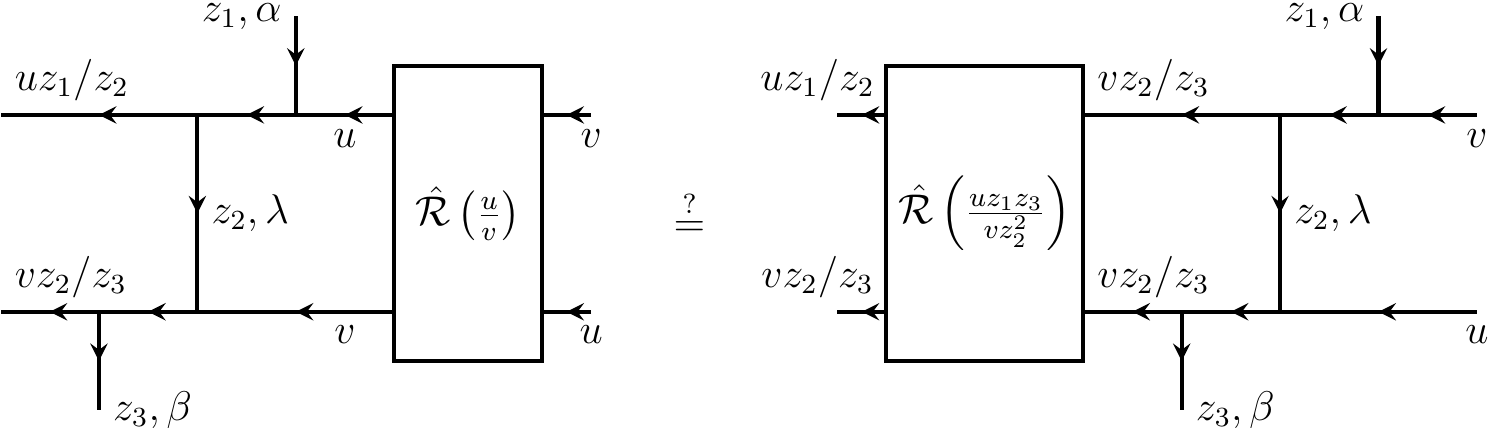}}
  \label{eq:63}
\end{equation}
\begin{multline}
\label{Rhor}
 \mathcal{V}^{\alpha}_{\beta} (u,v|z_1, z_2, z_3)   \hat{\cal R} \left( \frac{u}{v} \right) =
    \sum_\lambda
  \begin{array}{c}
    \mathcal{T}^{\alpha}_{\lambda}(u|z_1, z_2)\\
    \otimes\\
  \mathcal{T}^{\lambda}_{\beta}(v|z_2, z_3)
  \end{array} \hat{\cal R} \left( \frac{u}{v} \right) \stackrel{?}{=} \hat{\mathcal{R}} \left( \frac{uz_1z_3}{v z_2^2}
\right) \sum_\lambda
  \begin{array}{c}
    \mathcal{T}^{\alpha}_{\lambda}(v|z_1, \frac{z_1 z_3}{z_2})\\
    \otimes\\
  \mathcal{T}^{\lambda}_{\beta}(u|\frac{z_1 z_3}{z_2}, z_3)
\end{array} =\\
 =\hat{\mathcal{R}} \left( \frac{uz_1z_3}{v z_2^2} \right)
\mathcal{V}^{\alpha}_{\beta} \left(v, u\Big|z_1, \frac{z_1 z_3}{z_2}, z_3\right)
\end{multline}
These two matrices are dual in the sense that matrix elements of
$\hat{{\cal R}}$ coincide with those of ${\cal R}$ in another basis.
In the next section we will see that equalities~\eqref{eq:30}
and~\eqref{eq:63} are actually valid up to a nontrivial abelian
cocycle, the \emph{anomaly}.

\paragraph{5. Evaluation of the $R$-matrix and anomaly. Vertical
  representations.} As explained in \cite{Awata:2016mxc}, the
$R$-matrix is simple in the basis of generalized Macdonald
polynomials~\cite{Awata:2011dc,Morozov:2013rma,Mironov:2013oaa,Ohkubo:2014nqa,Zenkevich:2014lca},
where it only permutes the two Young diagrams, the spectral parameters
and the Fock spaces. In an appropriate normalization (without tilde in
the notations of~\cite{Awata:2016mxc}), the commutation relations read
as follows:
\begin{equation}
  \label{eq:4}
\boxed{{\cal R}_{\alpha
      \beta} \left( \frac{z_1}{z_2} \right)
    \mathcal{T}^{\alpha}_{\mu}\left(\frac{uz_2}{w_2}\Big|z_1,
      w_1\right) \mathcal{T}^{\beta}_{\nu}(u|z_2, w_2) =
   \mathcal{T}^{\beta}_{\nu}\left(\frac{uz_1}{w_1}\Big|z_2, w_2\right)
   \mathcal{T}^{\alpha}_{\mu}(u|z_1, w_1) {\cal R}_{\mu \nu} \left( \frac{w_1}{w_2} \right) E_{q,t} (z_1, w_1, z_2, w_2)}
\end{equation}
where $\mathcal{R}_{\alpha \beta}(x)$ are the diagonal elements of the
$R$-matrix in the basis of generalized Macdonald polynomials:
\begin{equation}
  \label{eq:13}
    \mathcal{R}^{\alpha \beta}_{\gamma\delta}(x) = \delta^{\alpha}_{\delta}
  \delta^{\beta}_{\gamma} \mathcal{R}_{ \beta \alpha}(x)
\end{equation}
given by
\begin{gather}
  \label{eq:5}
  {\cal R}_{\alpha \beta}(x) = \left(\frac{q}{t}
  \right)^{\frac{1}{2}(|\alpha|+|\beta|)} \frac{G_{\alpha \beta}(x)}{G_{\alpha
    \beta}\left( \frac{q}{t} x \right)}, \\
 G_{\alpha \beta}(x) = \prod_{(i,j)\in \alpha} \left( 1 - x
    q^{\alpha_i - j} t^{i - \beta^{\mathrm{T}}_j+1} \right)
  \prod_{(i,j)\in \beta} \left( 1 - x q^{-\beta_i + j-1} t^{-
      \alpha^{\mathrm{T}}_j+i} \right)=\nonumber\\
  = \exp \left[ \sum_{n \geq 1}
    \frac{1-t^n}{n
      (1-q^n)}  x^n \sum_{i,j}
    (q^{n(\alpha_i - \beta_j)} -1) t^{n(j-i)} \right]
    \label{eq:5p}
\end{gather}
and the ``anomalous'' factor $E$ reads
\begin{gather}
  \label{eq:8}
  E_{q,t} (z_1, w_1, z_2, w_2) =
  \frac{\Upsilon_{q,t}\left( \frac{q}{t}
      \Big| \frac{z_1}{z_2} \right)\Upsilon_{q,t}\left( 1
      \Big| \frac{w_1}{w_2} \right)}{\Upsilon_{q,t}\left( \sqrt{\frac{q}{t}}
      \Big| \frac{z_1}{w_2} \right)\Upsilon_{q,t}\left( \sqrt{\frac{q}{t}}
      \Big| \frac{w_1}{z_2} \right)},\\
  \Upsilon_{q,t}(\alpha|x) \stackrel{\mathrm{def}}{=} \exp \left\{
    \sum_{n\geq 1} \frac{1}{n} \frac{\alpha^n}{(1-q^n)(1-t^{-n})} (x^n
    - x^{-n}) \right\}
\end{gather}
Some elementary properties of the function
$\Upsilon_{q,t}$ are
\begin{gather}
  \Upsilon_{t^{-1},q^{-1}}(\alpha|x) = \Upsilon_{q,t} (\alpha|x),\notag\ \ \ \ \
  \Upsilon_{q,t}(\alpha|tx) = \frac{\Upsilon_{q,t}(\alpha|x)}{(\alpha
    tx;q)_{\infty} (\alpha x^{-1}; q)_{\infty}},\label{eq:14}
  \ \ \ \ \
  \Upsilon_{q,t}(\alpha|x^{-1}) = \frac{1}{\Upsilon_{q,t}
    (\alpha|x)}\notag
\end{gather}
The anomaly did not appear in our previous work~\cite{Awata:2016mxc}
because we normalized all the correlators there so that the vacuum
matrix elements were trivial on both sides of the $RTT$ relations. We
therefore studied only the dependence of the $R$-matrix and the $RTT$
relations on the Young diagram, in which the overall scalar factor
plays no role.

The $R$-matrix satisfies the usual identity
\begin{equation}
  \label{eq:12}
  {\cal R}_{\beta \alpha}(x^{-1}) = \frac{1}{{\cal R}_{\alpha \beta}(x)}.
\end{equation}
and turns into
unity in the unrefined limit, ${\cal R}_{\alpha
  \beta}(x)|_{t=q} = 1$.
Let us also mention that for certain values of the spectral parameters
the anomaly does not arise. For instance,
\begin{equation}
  \label{eq:11}
  E_{q,t}\left(z_1, \sqrt{\frac{t}{q}} z_1, z_2, \sqrt{\frac{t}{q}}  z_2\right) = E_{q,t}\left(z_1, \sqrt{\frac{q}{t}} z_1, z_2, \sqrt{\frac{q}{t}}  z_2\right)=
  1
\end{equation}
Notice that the values of $z_{1,2}$ and $w_{1,2}$ in Eq.~\eqref{eq:11}
coincide with the two special lines on the \emph{factorization loci}
of generalized Macdonald polynomials~\cite{MK}.

\paragraph{6. Vertical $R$-matrix from the universal DIM $R$-matrix.}
Let us show how the diagonal matrix $\mathcal{R}_{\alpha
  \beta}(x)$~\eqref{eq:5} can be obtained from the formula for the
universal DIM $R$-matrix considered in~\cite{FJMM}. As we have
mentioned in sec.~3 and Appendix E, the universal $R$-matrix in
principle depends on the choice of Borel subalgebra. The vertical
$R$-matrix is obtained from the ``vertical'' Borel subalgebra in the
DIM algebra. Similarly to the quantum affine algebra (again, see
Appendix E for a simplified example), the formula for the universal
DIM $R$-matrix in this case is given by:
\begin{equation}
  \label{eq:20}
  \mathcal{R}^{\perp} = \mathcal{K} \mathcal{R}_{\sim \delta}
\end{equation}
where
\begin{equation}
  \label{eq:54}
  \mathcal{K} = \left( \frac{q}{t} \right)^{\frac{1}{2} (c_{\perp} \otimes d_{\perp} + d_{\perp} \otimes c_{\perp})}, \qquad
  \mathcal{R}_{\sim \delta} = \exp \left[ - \sum_{n \geq 1} n (1-
    q^n)(1-t^{-n}) \left( 1 - \left( \frac{t}{q} \right)^n \right) h_n
    \otimes h_{-n}\right],
\end{equation}
$c_{\perp}$ is the ``vertical'' central charge (in our normalization
$c_{\perp}=1$ for the vertical Fock module) and $d_{\perp}$ is the
``vertical'' grading operator. $h_n$ in Eq.~\eqref{eq:54} are the
modes of the DIM generators $\psi^{\pm}(y)$:
\begin{equation}
  \label{eq:55}
  \psi^{\pm}(y) = \left( \frac{q}{t}\right) ^{\pm  \frac{1}{2}c_{\perp}} \exp \left[ \sum_{n\geq 1} (1-
    q^n)(1-t^{-n}) \left( 1 - \left( \frac{t}{q} \right)^n \right)
    h_{\pm n} y^{\mp n}  \right].
\end{equation}
Let us evaluate the $\mathcal{R}^{\perp}$ factor on the tensor product
of two vertical Fock modules with the spectral parameters $z_1$ and
$z_2$. The action of $\psi^{\pm}(y)$ is diagonal in the basis of
Macdonald polynomials~\cite{AFS}:
\begin{equation}
  \label{eq:56}
  \rho^{(0,1)}_z(\psi^{\pm}(y)) |M_{\lambda}, u \rangle =
  \left(\frac{q}{t} \right)^{\pm \frac{1}{2}} \prod_{i\geq 1}
  \frac{\left( 1 - \left(\frac{z}{y} x_i\right)^{\pm 1} \right) \left(
      1 - \left( \frac{t^2}{q}\frac{z}{y} x_i \right)^{\pm 1}
    \right)}{\left( 1 - \left( t
      \frac{z}{y} x_i \right)^{\pm 1} \right) \left( 1 -
      \left( \frac{t}{q} \frac{z}{y} x_i\right)^{\pm 1} \right)} |M_{\lambda}, u \rangle
\end{equation}
where $x_i = q^{\lambda_i} t^{-i}$. We thus find that
\begin{equation}
  \label{eq:58}
  \rho_{z}^{(0,1)}(h_{\pm n}) |M_{\lambda},z \rangle = \pm \frac{z^{\pm n} t^{\pm
      n}}{n (1 - q^{\pm n})} \sum_{i \geq 1} x_i^{\pm n} |M_{\lambda},z \rangle
\end{equation}
and
\begin{multline}
  \label{eq:59}
  \rho_{z_1}^{(0,1)} \otimes \rho_{z_2}^{(0,1)} \mathcal{R}^{\perp}
  |M_{\alpha},z_1 \rangle \otimes |M_{\beta},z_2 \rangle =\\
  = \left( \frac{q}{t} \right)^{\frac{1}{2}(|\alpha|+|\beta|)} \exp
  \left[ - \sum_{n \geq 1} \frac{(1-t^n) \left( 1 - \left( \frac{q}{t}
        \right)^n \right)}{n (1-q^n)} \left( \frac{z_1}{z_2} \right)^n
    \sum_{i,j} q^{n(\alpha_i - \beta_j)} t^{n(j-i)}
  \right]|M_{\alpha},z_1 \rangle \otimes |M_{\beta},z_2 \rangle =\\
  = f \left( \frac{z_1}{z_2} \right) \mathcal{R}_{\alpha \beta}\left(
    \frac{z_1}{z_2} \right) |M_{\alpha},z_1 \rangle \otimes
  |M_{\beta},z_2 \rangle.
\end{multline}
with $\mathcal{R}_{\alpha \beta}(x)$ given by Eq.~\eqref{eq:5}. The
prefactor $f(x)$ coincides with the one from~\cite{FJMM} (we therefore
have ${\cal R}_{\text{our}} =\bar R$ from~\cite{FJMM}):
\begin{equation}
  \label{eq:60}
  f(x) = \exp \left[ - \sum_{n \geq 1} \frac{(1-t^n) \left( 1 - \left(
          \frac{q}{t} \right)^n \right)}{n(1-q^n)} x^n \right].
\end{equation}
This calculation shows that the vertical $R$-matrix indeed can be
obtained from the universal DIM $R$-matrix for the vertical choice of
the Borel subalgebra. Moreover, this choice makes the $R$-matrix
diagonal in the basis of Macdonald polynomials, which are the spectral
duals of generalized Macdonald polynomials.

\paragraph{7. Evaluation of the $R$-matrix and anomaly. Horizontal
  representations.} The computation we have done in sec.~5 for the
vertical representations can be as well done for the horizontal
ones. To this end, consider the matrix element of the $RTT$ relations
in the horizontal channel~\eqref{Rhor} between two generalized
Macdonald polynomials:
\begin{multline}
  \label{eq:61}
  \left\langle M_{Y_1 Y_2}\left( u \frac{z_1}{z_2}, v \frac{z_2}{z_3}
    \right) \Big| \mathcal{V}^{\alpha}_{\beta} (u,v|z_1, z_2, z_3)
    \hat{\cal R} \left( \frac{u}{v} \right) \Big| M_{W_2 W_1} (v, u)
  \right\rangle \stackrel{?}{=}\\
  \stackrel{?}{=} \left\langle M_{Y_1 Y_2}\left( u \frac{z_1}{z_2}, v
      \frac{z_2}{z_3} \right) \Big|\hat{\mathcal{R}} \left(
      \frac{uz_1z_3}{v z_2^2} \right) \mathcal{V}^{\alpha}_{\beta}
    \left(v, u\Big|z_1, \frac{z_1 z_3}{z_2}, z_3\right) \Big| M_{W_2
      W_1} (v, u) \right\rangle
\end{multline}
The action of $\hat{\mathcal{R}}$ on the generalized Macdonald
polynomials is simple: up to a constant $\mathcal{R}_{Y_1 Y_2}$, it
just exchanges two variables, two spectral parameters and two
Young diagrams. The value of this constant can be derived by requiring
the Yang-Baxter equation and factorization of the $R$-matrices acting
on three representations to hold. The detailed calculation is
presented in Appendix B. We have
\begin{multline}
  \label{eq:62}
    \left\langle M_{Y_1 Y_2}\left( u \frac{z_1}{z_2}, v   \frac{z_2}{z_3}
    \right) \Big| \mathcal{V}^{\alpha}_{\beta} (u,v|z_1, z_2, z_3)
       \Big| M_{W_1 W_2} (u, v) \right\rangle \mathcal{R}_{W_1 W_2}
     \left( \frac{u}{v} \right) \stackrel{?}{=} \\
     \stackrel{?}{=} \mathcal{R}_{Y_1 Y_2} \left(
      \frac{uz_1z_3}{v z_2^2} \right) \left\langle M_{Y_2 Y_1}\left( v
      \frac{z_2}{z_3}, u \frac{z_1}{z_2} \right) \Big| \mathcal{V}^{\alpha}_{\beta}
    \left(v, u\Big|z_1, \frac{z_1 z_3}{z_2}, z_3\right) \Big| M_{W_2
      W_1} (v, u) \right\rangle
\end{multline}
The remaining matrix elements can be evaluated with the help of the matrix
model techniques. One can write them as the $q$-deformed Selberg
averages of the skew generalized Macdonald polynomials
(see~\cite{Morozov:2015xya} for the details):
\begin{multline}
  \label{eq:57}
  \left\langle M_{Y_1 Y_2}\left( u \frac{z_1}{z_2}, v \frac{z_2}{z_3}
    \right) \Big| \mathcal{V}^{\alpha}_{\beta} (u,v|z_1, z_2, z_3)
    \Big| M_{W_1 W_2} (u, v) \right\rangle =\\
  =   \left\langle \varnothing \Big| \mathcal{V}^{\alpha}_{\beta} (u,v|z_1, z_2, z_3)
    \Big| \varnothing \right\rangle \times\\
  \times \left\langle \sum_{Z_1, Z_2 } \left( \frac{q}{t} \right)^{|Z_1|+
      |Z_2|} \frac{1}{||M_{Z_1}||^2 ||M_{Z_2}||^2} M_{Y_1 Y_2/Z_1 Z_2}\left( u
      \frac{z_1}{z_2}, v \frac{z_2}{z_3} \right) M_{W_1 W_2/Z_1 Z_2}
    (u, v) \right\rangle_{q\text{-Selberg}}
\end{multline}
The $q$-Selberg average gives the bifundamental Nekrasov function
$z_{\mathrm{bif}}(\vec{Y}, \vec{W}) $ and some additional factors
$G_{Y_1 Y_2}$. The bifundamental part turns out to be the same on both
sides of Eq.~\eqref{eq:62}, and the additional factors exactly cancel
the $R$-matrices $\mathcal{R}_{Y_1 Y_2}$ (see~\cite{Awata:2016mxc} for
more details of this calculation). This proves the horizontal $RTT$
relations up to a scalar factor, $\langle\varnothing |
\mathcal{V}^{\alpha}_{\beta} (u,v|z_1, z_2, z_3) | \varnothing \rangle$.

This scalar factor comes from the \emph{vacuum} matrix elements of
$\mathcal{V}^{\alpha}_{\beta}$, which are different in the r.h.s.\ and
l.h.s. of Eq.~\eqref{eq:62}. For brevity, we provide the calculation
for $\alpha = \beta = \varnothing$, the general case being completely
analogous. Computing the matrix elements of $T$-operators explicitly
in this case, we get
\begin{multline}
  \label{eq:64}
  \left\langle \varnothing \Big| \mathcal{V}^{\varnothing}_{\varnothing} (u,v|z_1, z_2, z_3)
    \Big| \varnothing \right\rangle =  \begin{array}{c}
    \langle \varnothing |\mathcal{T}^{\varnothing}_{\lambda}(u|z_1,
  z_2)| \varnothing \rangle\\
    \otimes\\
  \langle \varnothing |\mathcal{T}^{\lambda}_{\varnothing}(v|z_2,
  z_3)| \varnothing \rangle
\end{array}=\\
= \sum_{\lambda} \left( - \frac{u z_3}{v z_2} \right)^{|\lambda|} M_{\lambda}^{(q,t)}\left(
  \frac{t^{\frac{n}{2}} \left( 1 - \left( \sqrt{\frac{t}{q}}
        \frac{z_1}{z_2} \right)^n \right)}{1- t^n} \right)
M_{\lambda^{\mathrm{T}}}^{(t,q)}\left(
  \frac{q^{\frac{n}{2}} \left( 1 - \left( \sqrt{\frac{q}{t}}
        \frac{z_2}{z_3} \right)^n \right)}{1- q^n} \right) =\\
=\exp
\left[  \sum_{n \geq 1} \frac{1}{(1-q^n)(1-t^{-n})} \left(
    \sqrt{\frac{q}{t}} \frac{uz_3}{vz_2} \right)^n \left( 1 - \left(
      \sqrt{\frac{t}{q}} \frac{z_2}{z_3}  \right)^n - \left(
      \sqrt{\frac{q}{t}} \frac{z_1}{z_2}\right)^n + \left( \frac{z_1}{z_3} \right)^n \right) \right]
\end{multline}
Dividing the vacuum matrix elements in the left and right hand sides
of the $RTT$ relations \eqref{eq:62} we get precisely the anomaly
coefficient:
\begin{equation}
  \label{eq:65}
  \frac{ \left\langle \varnothing \Big| \mathcal{V}^{\varnothing}_{\varnothing} (u,v|z_1, z_2, z_3)
    \Big| \varnothing \right\rangle}{\left\langle \varnothing \Big|\mathcal{V}^{\varnothing}_{\varnothing} \left(v, u\Big|z_1, \frac{z_1
      z_3}{z_2}, z_3\right) \Big| \varnothing \right\rangle} = E_{q,t}
(z_1^{\perp}, w_1^{\perp}, z_2^{\perp}, w_2^{\perp})
\end{equation}
where the ``perpendicular'' variables can be read off from the
pictures in Eq.~\eqref{eq:63}:
\begin{gather}
  \label{eq:66}
  z_1^{\perp} = u,\qquad z_2^{\perp} = v,\\
  w_1^{\perp} = u \frac{z_1}{z_2}, \qquad w_2^{\perp} = v \frac{z_2}{z_3}.
\end{gather}
Finally, we have the horizontal $RTT$ relation:
\begin{equation}
  \label{eq:67}
  \boxed{   \mathcal{V}^{\alpha}_{\beta} (u,v|z_1, z_2, z_3)
    \hat{\cal R} \left( \frac{u}{v} \right)
    = \hat{\mathcal{R}} \left(
      \frac{uz_1z_3}{v z_2^2} \right) \mathcal{V}^{\alpha}_{\beta}
    \left(v, u\Big|z_1, \frac{z_1 z_3}{z_2}, z_3\right) E_{q,t} \left(u, u
    \frac{z_1}{z_2}, v, v \frac{z_2}{z_3}\right)}
\end{equation}
It is completely equivalent to the vertical $RTT$
relations~\eqref{eq:4}, which are obtained by the action of the
spectral duality.

\paragraph{8. Anomaly cancellation and group element.} As the $RTT$
relations are anomalous, the $T$-operator is not the true group
element of the DIM algebra. However, it is possible to build such an
element, which would satisfy the usual $RTT$ relations \emph{without}
the anomaly. In fact, one can remove the anomaly factor by the recipe
somewhat similar to 't Hooft's anomaly matching. In this approach the
gauge anomalies of the system of interest are cancelled by introducing
an auxiliary weakly coupled sector charged under the same gauge
group. This new sector is engineered so as to produce the anomaly
exactly \emph{opposite} to that of the original system. Hence, the
total system becomes non-anomalous.

We construct our auxiliary system as follows (see Appendix C for
details). The function $E_{q,t}$ is nothing but the four-point free
field correlator of the form
\begin{equation}
  \label{eq:10}
  E_{q,t}(z_1, w_1, z_2, w_2) \sim \langle \varnothing | \Psi^{*}_{\varnothing}(w_1)
  \Psi^{\varnothing}(z_1)  \Psi^{*}_{\varnothing}(w_2)
  \Psi^{\varnothing}(z_2) |  \varnothing \rangle
\end{equation}
and the function $\Upsilon_{q,t}$ plays the role of the pair
correlator of $\Psi$-fields. We introduce \emph{auxiliary} Fock
spaces living on each \emph{horizontal} leg of the toric diagram and
multiply the intertwiners with auxiliary operators
$\widetilde{\Psi}^{\varnothing}$ and
$\widetilde{\Psi}^{*}_{\varnothing}$ acting on these extra Fock spaces
as free field exponentials:
\begin{gather}
\Psi^\lambda(z) \to e^{i\tilde\Phi(z)}\Psi^\lambda(z)\\
\Psi^*_\lambda(z)\to e^{-i{\tilde\Phi^*}(z)}\Psi^*_\lambda(z)
\end{gather}
Naturally these extra operators commute
with the original intertwiners as well as with the whole DIM
algebra. However, they do not commute among themselves, and, as we
show in Appendix C, produce the \emph{inverse} of the anomaly factor
$E_{q,t}$, thus cancelling the total anomaly.

Unfortunately, this recipe works only for the \emph{vertical} $RTT$
relations (sec.~5). To deal with the anomaly in the \emph{horizontal}
$RTT$ relations (sec.~7), one adds one more Fock space living on
the vertical legs as well as some horizontal legs and multiplies the
intertwiners with extra factors $\widetilde{\widetilde{\Psi}}{}^{\mu}$
and $\widetilde{\widetilde{\Psi}}{}^{*}_{\mu}$ (again we refer to
Appendix C for details). With this somewhat contrived construction one
indeed can cancel the anomalies in all the $RTT$ relations, though the
price to pay are extra complications.

The resulting $T$-operator plays the role of the DIM group
element satisfying the usual non-anomalous $RTT$ relations. Let us
mention that one obtains in this way the $R$-matrix that coincides
with the normalized $\bar R$-matrix from \cite{FJMM}, the resulting
$T$-operator should be associated with the normalized $\bar
T$-operator from~\cite{FJMM}.

\paragraph{9. The ``vacuum'' case.}
When two of the Young-diagram indices are empty, the $R$-matrix ${\cal
  R}_{\emptyset\emptyset}^{\gamma\delta}$ trivializes, but the
corresponding $T$-operators with empty vertical legs still do not
commute due to the anomaly:
\begin{equation}
   \mathcal{T}^{\varnothing}_{\varnothing}\left(\frac{uz_2}{w_2}\Big|z_1,
     w_1\right) \mathcal{T}^{\varnothing}_{\varnothing}(u|z_2, w_2) =
   \mathcal{T}^{\varnothing}_{\varnothing}\left(\frac{uz_1}{w_1}\Big|z_2,
     w_2\right) \mathcal{T}^{\varnothing}_{\varnothing}(u|v_1, u_1)    E_{q,t} (z_1, w_1, z_2, w_2)
\end{equation}
Then, (\ref{Rvert}) implies that the operators
$\mathcal{V}^{\varnothing}_{\varnothing}$ defined in
Eq.~\eqref{eq:15}, also commute in a similar way:
\begin{multline}
  \mathcal{V}^{\varnothing}_{\varnothing}\left(
    \frac{uz_2}{w_2},\frac{v w_2}{y_2} \Big| z_1, w_1, y_1\right)
  \mathcal{V}^{\varnothing}_{\varnothing}(u,v|z_2, w_2, y_2) =\\
  =   \mathcal{V}^{\varnothing}_{\varnothing}\left(
    \frac{uz_1}{w_1},\frac{v w_1}{y_1}\Big| z_2, w_2, y_2\right) \mathcal{V}^{\varnothing}_{\varnothing}(u,v|z_1, w_1, y_1)
  E_{q,t} (z_1, w_1, z_2, w_2) E_{q,t}(w_1, y_1, w_2, y_2)
  \label{eq:35}
\end{multline}
This is a trivial implication of generic quantum group theory to the DIM
algebra. A direct consequence is a drastic simplification of the
modular properties.  It is not directly seen at the level of
ordinary $5d$ Nekrasov functions, because these are only coefficients
of the formal series, while the $RTT$ relations describe properties of the
full Nekrasov functions ($5d$ conformal blocks) obtained by
appropriate summation of the series.

\bigskip

The $4d$ limit of the function $E_{q,t}$ is very simple: it just
becomes a combination of powers (as we mention in Appendix D, it is important to scale the vertical parameters
appropriately):
\begin{equation}
  \label{eq:9}
  E_{q,t}(z_1,w_1,z_2,w_2)|_{q\to 1, t= q^{\beta}, z_1/w_1 =
    q^{\alpha_1}, z_2/w_2 =
    q^{\alpha_2}} \to  (1 - z_1)^{\alpha_1
    \alpha_2}(1-z_2)^{\alpha_1 \alpha_2}
\end{equation}
Notice that the \emph{ratio} of $z$ and $w$ in each vertex operator
scales as $q^{\alpha}$, where $\alpha$ corresponds to Liouville
momentum of the field. The $4d$ anomaly function~\eqref{eq:9} is
responsible for the commutation relations of the \emph{unscreened} CFT
vertex operators.

\bigskip

\paragraph{10. Integrals of motion and $6d$ theories.} Usually, if there are
the $RTT$ relations, the integrals of motion immediately follow. To
this end, one simply takes the trace in the appropriate spaces, which provides the commutativity of the transfer matrices $\tr
\mathcal{T}(z)$. However, the anomaly introduces additional
complications.

One can take the trace over the vertical lines in Eq.~\eqref{eq:30}
and additionally shift the spectral parameters of all the vertical
representations by an arbitrary parameter $Q_B$, i.e. set $w_{1,2} = Q_B z_{1,2}$. Then, the
commutation relation for the traces of the $T$-operators
follows from (\ref{eq:4}) and looks as follows (with an arbitrary weight parameter $Q_F$):
\begin{multline}
  \label{eq:51}
  \sum_{\alpha, \beta} Q_F^{|\alpha| + |\beta|}
  \mathcal{T}^{\alpha}_{\alpha} (uQ_B^{-1} | z_1, Q_B z_1)
  \mathcal{T}^{\beta}_{\beta} (u| z_2, Q_B z_2) =\\
  = \sum_{\alpha, \beta} Q_F^{|\alpha| + |\beta|}
  \mathcal{T}^{\beta}_{\beta} (uQ_B^{-1}| z_2, Q_B z_2)
  \mathcal{T}^{\alpha}_{\alpha} (u| z_1, Q_B z_1) E_{q,t} (z_1, Q_B
  z_1, z_2, Q_B z_2)
\end{multline}
One can see, that the operators $t_{Q_B, Q_F}(u|z) = \sum_{\alpha}
Q_F^{|\alpha|} \mathcal{T}^{\alpha}_{\alpha}(u|z, Q_B z)$ do not
commute because of the anomaly factor. However, due to the
identities~\eqref{eq:11} for particular values $Q_B = \left(
  \frac{q}{t}\right)^{\pm \frac{1}{2}}$ the traces are in fact
commutative.

To get a gauge theory interpretation of these results, we take the
vacuum matrix element of Eq.~\eqref{eq:51}:
\begin{gather}
  \label{eq:52}
  Z(z_1, z_2, Q_B, Q_F) = \langle \varnothing |
  \sum_{\beta} Q_F^{|\alpha|}  \mathcal{T}^{\alpha}_{\alpha}
  (uQ_B^{-1} | z_2, Qz_2) \sum_{\beta}
  Q_F^{|\beta|} \mathcal{T}^{\beta}_{\beta} (u| z_1, Q_Bz_1) | \varnothing
  \rangle\\
  Z(z_2, z_1, Q_B, Q_F) = Z(z_1, z_2, Q_B, Q_F) E_{q,t} (z_1, Q_B z_1, z_2, Q_B
  z_2)\label{eq:53}
\end{gather}

The partition function $Z(z_1, z_2, Q_B,Q_F)$ corresponding to the traces
of $T$-operators has two different gauge theory interpretations
connected by the spectral duality. One of them is the $5d$ $U(2)$
adjoint theory, where $Q_B$ plays the role of the adjoint mass,
$z_{1,2}$ are Coulomb moduli and $Q_F$ is the coupling
constant. Eq.~\eqref{eq:53} can be understood as an anomaly in the
Weyl group of the gauge group, which makes a transformation $z_1
\leftrightarrow z_2$ nontrivial (though in a controllable way). Let us
mention that the anomaly actually arises from the $U(1)$ factor, and
this is the reason why its contribution in Eq.~\eqref{eq:53} is
factorized.

The second interpretation of $Z(z_1, z_2, Q_B, Q_F)$ is the $6d$
$U(1)$ theory with one fundamental and one antifundamental
hypermultiplet. In this setting, $Q_B$ is the exponentiated radius of
the sixth dimension, $\Lambda = \frac{z_1}{z_2}$ is the exponentiated
coupling constant and $Q_F$ controls the masses of the
hypermultiplets, which are constrained to add up to zero in $6d$
theory to cancel the gauge anomaly. Eq.~\eqref{eq:53} then describes
the \emph{braiding} properties of the partition function under the
transformations $\Lambda \leftrightarrow \Lambda^{-1}$.

Let us also mention that the traces of $T$-operators are related to
the spectrum of certain integrable $2d$ field theories, like the
difference version of the ILW hierarchy~\cite{ILW}. More concretely,
the traces are intertwining operators of the \emph{elliptic} DIM
currents, which are known to contain the ILW Hamiltonians as zero
modes. This fact points out a remarkable connection between the affine
and elliptic systems. We will elaborate on this subject in a future
work.

\paragraph{11. Anomaly as the origin of braiding of conformal blocks.}
The anomaly in the $RTT$ relations manifests itself as a nontrivial
commutation relation for the screened vertex operators of $q$-Virasoro
algebra. To see this connection, one should recall how the conformal
block arises from correlators of topological vertices. The spectral
parameters of the vertical representations correspond to the positions
of the vertex operator insertions. The $R$-matrix exchanges $z_1
\leftrightarrow z_2$, which in the language of CFT means exchanging
positions of the two vertex operators and their dimensions, i.e.\
making the braiding transformation.

The conformal block is the vacuum matrix element of the combination of
$T$-operators. For example, for the four-point Virasoro block one
has:
\begin{equation}
  \label{eq:34}
  B_4 = \langle \varnothing | \mathcal{V}^{\varnothing}_{\varnothing}\left(
    \frac{uz_1}{w_1},\frac{v w_1}{y_1} \Big| z_2, w_2, y_2\right)
  \mathcal{V}^{\varnothing}_{\varnothing}(u,v|z_1, w_1, y_1) |\varnothing\rangle
\end{equation}
Eq.~\eqref{eq:35} demonstrates that if there was no anomaly, the
braiding transformation would act trivially on the block. Thus, one
naturally relates the anomaly function with the braiding kernel in
$q$-deformed CFT. Development along this line will be reported
elsewhere.

\paragraph{12. Knot invariants and $R$-matrix.} Let us discuss the
construction of knot and link invariants from an $R$-matrix
\cite{RT}. Each knot can be represented (not uniquely) as a closure of
a braid. To each crossing of strands, one associates the $R$-matrix
and the closure is given by the (quantum) trace over the
representation space of the $R$-matrix. For this
construction~\cite{RTmod} to work, the $R$-matrix should satisfy the
braid group relations \emph{and} the trace should be well defined.

As we have seen, the $R$-matrix~\eqref{eq:31} satisfies the braid
group relations~\eqref{eq:6}. The ordering of the spectral parameters
might seem strange, since it does not agree with the ordering of the
tensor indices. However, one can look at the spectral parameter
residing on a given strand simply as an additional parameter of the
representation (which it really is, since the representations in
question are evaluation representations of the quantum affine
algebras). The trace on the evaluation representations is also easy to
define.

One can wonder why then there are no knot invariants associated with
the $R$-matrices with spectral parameters? The answer lies in the
special property of the $R$-matrices, which in fact can be linked to
its analytic structure. Consider the $R$-matrix acting on two strands
with spectral parameters $z_1$ and $z_2$ respectively. In our case,
the definition of the \emph{opposite} $R$-matrix $R^{*}$, which
corresponds to the crossing opposite to that of $R$ is
\begin{equation}
  \label{eq:2}
  R^{*}(z_1, z_2) = R^{-1}(z_2, z_1)
\end{equation}
This is a simple consequence of the second Reidemeister move (see,
e.g., \cite{Reid}).  However, the usual $R$-matrix with spectral
parameter has a very special property~(\ref{eq:12}), that is,
\begin{equation}
  \label{eq:32}
  R^{-1}(z_2, z_1) = R(z_1, z_2)
\end{equation}
Combining Eqs.~\eqref{eq:2} and~\eqref{eq:32}, one gets a
remarkable result
\begin{equation}
  \label{eq:33}
  R^{*}(z_1, z_2) = R(z_1, z_2)
\end{equation}
Thus, the $R$-matrix does not depend on the way the strands are
crossed: the opposite crossings produce the same result\footnote{Recently this issue was also raised in \cite{Witnew} with the emphasis on an alternative approach due to \cite{Cas}.}. As a
simple example, the Hopf link and two unknots have the same trivial
invariant, since $R^2 = R R^{-1} = 1$.

There is one more way to understand the relation~\eqref{eq:33}. The
$R$-matrix actually depends on the \emph{ratio} of the two spectral
parameters. Thus, the two sides of Eq.~\eqref{eq:33} are series in
different variables: $\frac{z_1}{z_2}$ and $\frac{z_2}{z_1}$. One of
them is valid in the vicinity of $\frac{z_1}{z_2} = 0$ and the other
one in the neighbourhood of $\frac{z_1}{z_2} = \infty$. The statement
of Eq.~\eqref{eq:33} is that these series actually agree with each
other, i.e.\ that $R^{-1}$ is the analytic continuation of $R$ from
small to large values of the spectral parameters. It can in fact
happen that the $R$-matrix contains extra singularities so that the
analytic continuation does not work in a naive way. If this is the
case, Eq.~\eqref{eq:33} will no longer hold and it is in principle
possible to obtain a \emph{nontrivial} knot invariant from such an
$R$-matrix.  The paper~\cite{Aganagic-Okounkov} hints that the
analytic continuation in some cases is in fact nontrivial.

\paragraph{Conclusion.}  We have studied the intertwining properties
of two $T$-operators, which are the liftings of DF-screened CFT vertex
operators to network matrix models. Correlators of $\mathcal{T}$
satisfy $qq$-character equations (the lift of the matrix
model/$\beta$-ensemble Virasoro constraints). As noticed
in~\cite{Awata:2016mxc}, these operators satisfy the $RTT$ relations with
the DIM-algebra $R$-matrix, which we explicitly calculate (in
the simplest representations) in both the horizontal and vertical
channels.  However, the $RTT$ relations turn out to hold only modulo
an abelian anomaly factor (which is the same in both channels, in full
accordance with expectations from the spectral duality
of~\cite{specdu}).  Algebraically, the anomaly means that our (network
model) $T$-operator is not quite a true group element. However, since
the anomaly is pure abelian, it can be easily eliminated
by multiplying the $T$-operator with additional factors made from extra free fields, the mechanism being similar to the anomaly matching condition in gauge theory.
However, physically this anomaly seems to be
absolutely relevant, because it is needed to reproduce the
non-commutative operator product expansion (OPE) of CFT vertex
operators.

Our results are consistent with the previous calculations of universal
$R$-matrix in \cite{FJMM} and non-trivially extend them from
vertical to horizontal channel, where the generalized Macdonald
polynomial technique of~\cite{Morozov:2013rma, Ohkubo:2014nqa,
  Zenkevich:2014lca} is needed and successfully applied.  We also
explain, why the emerging DIM $R$-matrix can not be used in
knot theory calculations: this is not because it depends on a spectral
parameter, but because of a peculiar symmetry~\eqref{eq:2} in this
dependence, which, however, can disappear in more general
representations of DIM (currently described only in sophisticated
combinatorial terms of 3d partitions). This possibility adds to the
motivations for further investigations of the $RTT$ relations, which
can require a development of the non-abelian free field techniques for
the toroidal algebras, similar to those from \cite{GMMOS} for the
affine ones.

\paragraph{Appendix A: Details of the free field formalism.}
In the horizontal representation, elements of the DIM algebra act as
the exponentiated currents $\eta(z)$ and $\xi(z)$ built from the free
bosonic field $\phi(z)$. For vertical representations, the DIM action
on the basis of Macdonald polynomials $|M_{\lambda}^{(q,t)}\rangle$ is
realized combinatorially. $\Psi^{\lambda}$ and $\Psi^{*}_{\lambda}$
defined in Eq.~\eqref{eq:1} are partial \emph{matrix elements} of the
intertwiners obtained by plugging the vector
$|M_{\lambda}^{(q,t)}\rangle$ from the vertical representation into
the intertwiner $\Psi$ (or $\Psi^{*}$) acting in the tensor product of
the horizontal and vertical representations, $\Psi^{\lambda} = \Psi
|M_{\lambda}\rangle^{\mathrm{vert}}\otimes\, \cdot\,$. The concrete
expressions for the intertwiners are given by \begin{align}
  \label{eq:23}
  \parbox{3cm}{\includegraphics[width=3cm]{figures/Psi-crop}} &= (-zu)^{|\lambda|}(-z)^{-(N+1)|\lambda|}f_{\lambda}^{-N-1}
  \frac{q^{n(\lambda^{\mathrm{T}})}}{C_{\lambda}}
  :\Psi_{\varnothing} (z) \prod_{(i,j)\in \lambda} \eta(z q^{j-1} t^{1-i}):
 \\ \nn \\
  \parbox{3cm}{\includegraphics[width=3cm]{figures/Psistar-crop}} &=
  (q^{-1}u)^{-|\lambda|} (-w)^{N|\lambda|} f_{\lambda}^N
  \frac{q^{n(\lambda^{\mathrm{T}})}}{C_{\lambda}}
  :\Psi^{*}_{\varnothing}(w) \prod_{(i,j)\in \lambda} \xi(w q^{j-1}
  t^{1-i}):\label{eq:24}
\end{align}
where the superscript T means the transposed Young diagram.  The
normal ordered combinations of operators in the vertices act in the
``horizontal'' Fock space and are defined in terms of the $q$-deformed
free field
\begin{gather}
  \label{eq:21}
  \phi(z) =  \phi_{-}(z) + \phi_{+}(z) = \sum_{n \geq 1} \frac{1}{n} a_{-n} z^n
  -  \sum_{n \geq 1} \frac{1}{n} a_n z^{-n}   \nn\\
    \phi^{*}(z) =  \phi_{-}\left( \sqrt{\frac{t}{q}} z\right) +
    \phi_{+}\left( \sqrt{\frac{q}{t}} z\right) = \sum_{n \geq 1} \frac{1}{n} a^{*}_{-n} z^n
  -  \sum_{n \geq 1} \frac{1}{n} a^{*}_n z^{-n} = \sum_{n \geq 1}
  \frac{1}{n} \left( \frac{t}{q} \right)^{\frac{n}{2}} a_{-n} z^n
  -  \sum_{n \geq 1} \frac{1}{n} \left( \frac{t}{q} \right)^{\frac{n}{2}} a_n z^{-n}   \nn\\
  \left[  a_n, a_m\right] = n \frac{1 - q^{|n|}}{1 - t^{|n|}}
  \delta_{m+n,0}, \qquad \qquad \left[  a_n^{*}, a_m^{*}\right] = n
  \left( \frac{t}{q} \right)^{|n|} \frac{1 - q^{|n|}}{1 - t^{|n|}} = n \frac{1 - q^{-|n|}}{1 - t^{-|n|}}
  \delta_{m+n,0}
\end{gather}
by the following formulas:
\begin{gather}
  \eta(z) = :e^{\phi(z) - \phi(z/t)}: \nn \\
  \xi(w) = :e^{-\phi^{*}(w) + \phi^{*}(w/t)}: 
  \nn\\
  \Psi_{\varnothing}(z) = :\prod_{k \geq 0} e^{-\phi(q^k z)}:\nn\\
  \Psi^{*}_{\varnothing}(w) = :\prod_{k \geq 0} e^{\phi^{*}(q^k w)} 
  :  \label{eq:16}
\end{gather}
The infinite products here should
  be carefully regularized, so that the resulting operators make
  sense. We do not concentrate on this subtlety and only mention that
  the regularization indeed can be performed.

These vertices are invariant under the simultaneous exchange
of $\lambda \to \lambda^{\mathrm{T}}$ and $q \to t^{-1}$. The spectral
parameter of the representation can also be understood as the
eigenvalue of the zero mode of the free field $\phi(z)$. The framing
factors are
\begin{gather}
  f_{\lambda}(q,t) = (-1)^{|\lambda|} q^{n(\lambda^{\mathrm{T}}) +
    \frac{|\lambda|}{2}} t^{-n(\lambda) - \frac{|\lambda|}{2}} \nn \\
  C_{\lambda} (q,t)= \prod_{(i,j)\in \lambda} \left( 1 - q^{\lambda_i - j}
    t^{\lambda^{\mathrm{T}}_j - i + 1} \right) \nn \\
  n(\lambda) = \sum_{(i,j)\in \lambda} (i-1).
  \label{def}
\end{gather}
Using the coproduct of the DIM algebra, one can take tensor products of
several parallel horizontal representations. In this tensor product acts the $W_m$ algebra, which can be thought of as a subalgebra of
DIM. More concretely~\cite{FHSSY-kernel}, from the generators of the
DIM algebra one can build a dressed current $t(z)$, which in the
tensor product of $m$ Fock representations acts as
$\Delta^{m-1}(t(z))$ and produces the energy-momentum tensor of the
$W_m$-algebra. Higher spin currents can be obtained by the Miura
transform.

\paragraph{Appendix B: DIM $R$-matrix and the integral form of
  generalized Macdonald polynomial.} Let us rederive the formulas of
sec.~7 directly within the framework of the generalized Macdonald
polynomials.  Recall that the DIM $R$-matrix acts on the generalized
Macdonald polynomials as \cite{Awata:2011dc, Morozov:2013rma,
  Mironov:2013oaa, Ohkubo:2014nqa}
\begin{equation} {\cal \hat{R}} \left( M_{\lambda\mu}(u_1,u_2 | q,t |
    p^{(1)}, p^{(2)}) \right) = \mathcal{R}_{\lambda\mu}
  M_{\lambda\mu}^{\rm op} (u_1,u_2 | q,t | p^{(1)},
  p^{(2)})~, \label{Rdef}
\end{equation}
where
\begin{equation}
M_{\lambda\mu}^{\rm op} (u_1,u_2 | q,t | p^{(1)},  p^{(2)}) :=
M_{\mu\lambda}(u_2,u_1 | q,t | p^{(2)},  p^{(1)}).
\end{equation}
Our initial normalization of the generalized Macdonald polynomial is
\begin{equation}
M_{\lambda\mu}(u_1,u_2 | q,t | p^{(1)},  p^{(2)}) = m_\lambda(p^{(1)}) m_\mu(p^{(2)}) + \cdots,
\end{equation}
where $m_\lambda$ is the monomial symmetric function.  When we compute
the $R$-matrix by \eqref{Rdef}, it is important to fix the
proportionality constant $\mathcal{R}_{\lambda\mu} =
\mathcal{R}_{\lambda\mu}(u_1, u_2 | q,t )$, which can be obtained by
the method of \cite[Appendix A]{Awata:2016mxc}.  By computing
$\mathcal{R}_{\lambda\mu}$ explicitly for lower levels $(|\lambda| + |\mu| \leq
3)$, we arrive at the following formula (see \eqref{eq:5})
\begin{equation}\label{eq: kconjecture}
\mathcal{R}_{\lambda\mu}
= \left( \frac{q}{t} \right)^{\frac{1}{2}(|\lambda| + |\mu|)}
\frac{G_{\lambda\mu}\left( \frac{u_1}{u_2} ; q,t \right)} {G_{\lambda\mu}\left( \frac{qu_1}{tu_2} ; q,t \right)}
= \left( \frac{t}{q} \right)^{\frac{1}{2}(|\lambda| + |\mu|)}
\frac{G_{\mu\lambda}\left( \frac{qu_2}{tu_1} ; q,t \right)} {G_{\mu\lambda}\left( \frac{u_2}{u_1} ; q,t \right)}~,
\end{equation}
where $G_{\lambda\mu}(Q; q,t)$ defined in (\ref{eq:5p}) is the factor appearing in the vector multiplet part of
the Nekrasov partition function. The second equality follows from
the formula (\cite[Eq.(2.34)]{Awata:2008ed}, \cite[Eq.(102)]{Morozov:2015xya})
\begin{equation}
G_{\lambda\mu}\left( \sqrt{\frac{q}{t}} Q ; q,t \right)
= G_{\mu\lambda}\left( \sqrt{\frac{q}{t}} Q^{-1} ; q,t \right)
Q^{|\lambda| + |\mu|} \frac{f_\lambda(q,t)}{f_\mu(q,t)},
\end{equation}
where $f_\lambda(q,t)$ is the framing factor (\ref{def}), \cite{Taki:2007dh}.
Employing the same formula, we also obtain
\begin{equation}
  \mathcal{R}_{\lambda\mu} = \beta_{\lambda\mu} \frac{N_{\mu\lambda} \left( u_2, u_1 ; q,t \right)}
  {N_{\lambda\mu} \left( u_1, u_2 ; q,t \right)}~, \label{kformula}
\end{equation}
where
\begin{equation}
\beta_{\lambda\mu}(u_1, u_2 | q,t )
 = \left( \frac{u_2}{u_1} \right)^{|\lambda| + |\mu|} \frac{f_\mu(q,t)}{f_\lambda(q,t)}~,  \label{betadef}
\end{equation}
and
\begin{equation}
N_{\lambda\mu} (u_1, u_2; q,t) := G_{\mu\lambda} (\frac{u_2}{u_1}; q,t) C_\lambda(q,t) C_\mu(q,t)
\end{equation}
is the normalization of the generalized Macdonald polynomial in \cite{Zenkevich:2014lca} ($C_\lambda$ is defined in (\ref{def})).
In \cite{Awata:2016mxc}, we normalized $\beta_{\lambda\mu} \equiv 1$ (see Eq.(74)).
Here in \eqref{betadef} we restore the full normalization.
Note that \eqref{eq: kconjecture} satisfies the consistency condition
\begin{equation}
  \mathcal{R}_{\lambda\mu}(u_1, u_2 | q,t) \mathcal{R}_{\mu\lambda} (u_2, u_1 | q,t) =1,
\end{equation}
which means $P_{12} {\cal R}(u_2, u_1)  P_{12} {\cal R} (u_1, u_2)=
{\cal R}^{\rm op}(u_1, u_2){\cal R} (u_1, u_2) =1$,
where $P_{12}$ exchanges $p^{(1)}$ and $p^{(2)}$;
\begin{equation}
P_{12} M_{\lambda\mu}(u_1,u_2 | q,t | p^{(2)},  p^{(1)})  = M_{\lambda\mu}(u_1,u_2 | q,t | p^{(1)},  p^{(2)})~.
\end{equation}

One can make $\mathcal{R}_{AB}$ in \eqref{Rdef} trivial in the
following way. Let us introduce
\begin{equation}\label{eq:integral normalization}
\widetilde{N}_{\lambda\mu} (u_1,u_2, q,t) := (-1)^{|\lambda|} u_1^{|\lambda| + |\mu|} (q/t)^{\frac{|\mu|}{2}}
q^{-n(\mu^T) - |\mu|} t^{-n(\lambda)- |\lambda|}N_{\lambda\mu} (u_1, u_2, q,t),
\end{equation}
which satisfies
\begin{equation}
  \frac{\widetilde{N}_{\mu\lambda} \left( u_2, u_1; q,t \right)}
  {\widetilde{N}_{\lambda\mu} \left( u_1, u_2 ; q,t \right)} = \mathcal{R}_{\lambda\mu}.
\end{equation}
Then, in the {\it special} normalization
\begin{equation}
K_{\lambda\mu}(u_1,u_2 | q,t | p^{(1)},  p^{(2)})  :=
\widetilde{N}_{\lambda\mu} \left(\frac{u_1}{u_2}, q,t\right) M_{\lambda\mu} (u_1,u_2 | q,t | p^{(1)},  p^{(2)}),
\end{equation}
one gets
\begin{equation}
{\cal \hat{R}} \left( K_{\lambda\mu}(u_1,u_2 | q,t | p^{(1)},  p^{(2)}) \right)
= K_{\lambda\mu}^{\rm op} (u_1,u_2 | q,t | p^{(1)},  p^{(2)}).
\end{equation}
We have explicitly checked for lower levels that $K_{\lambda\mu}$ agrees with the integral form of the generalized Macdonald
polynomials \cite{Awata:2011dc} defined by the PBW type basis of the DIM algebra:
\begin{equation}
\ket{X_{\vl}}= \Xo_{-\lo_1} \Xo_{-\lo_2} \cdots \Xt_{-\lt_1} \Xt_{-\lt_2}.  \cdots \ket{0},
\end{equation}
with
\begin{eqnarray}
X^{(1)} (z) &:=& (\rho_{u_1} \otimes \rho_{u_2} ) (\Delta(x^{+}(z)) = u_1 \Lambda_1(z) + u_2  \Lambda_2(z) ~, \\
X^{(2)} (z) &:=& u_1 u_2 :  \Lambda_1(z) \Lambda_2((q/t) z) :.
\end{eqnarray}
The representation matrix ${\cal R}_{\vl, \vm}$ in the basis of the integral form can be expressed as
\begin{equation}\label{eq:formula of R Matrix}
{\cal R}_{\vl, \vm}{=}
\frac{1}{\braket{K^*_{\vm} }{K_{\vm}}}
\braket{K^*_{\vm}}{K^{\mathrm{op}}_{\vl}},
\end{equation}
where $K^*_{\vl}$ is the dual basis with respect to the inner product
defined in terms of the power sum polynomial $p_{\vec{\lambda}}$;
\begin{equation}
\left\langle p_{\vec{\lambda}}| p_{\vec{\mu}} \right\rangle
:=\delta_{\vec{\lambda}, \vec{\mu}} \prod_{i=1}^{N} z_{\lambda^{(i)}}
 \prod_{k=1}^{\ell (\lambda^{(i)})}\frac{1-q^{\lambda_k^{(i)}}}{1-t^{\lambda_k^{(i)}}} , \qquad
z_{\lambda^{(i)}} \mathbin{:=} \prod_{k \geq 1}k^{m_k} \, m_k !.
\end{equation}
Let us introduce the transition matrix from the (tensor) product of the power sum
polynomials\footnote{One may use any basis in the space of symmetric polynomials.}
to the integral form of the generalized Macdonald polynomials
\begin{equation}
K_{\vl} (u_1,u_2 | q,t | p^{(1)},  p^{(2)}) =\sum_{\vm}
\mathcal{A}_{\vl,\vm}(u_1,u_2 | q,t) p_{\vec{\mu}}
\end{equation}
and its opposite version
\begin{equation}
\mathcal{A}^{\mathrm{op}}_{(\lo,\lt), (\mo, \mt)}(u_1, u_2)
:=\mathcal{A}_{(\lt,\lo), (\mt, \mo)}(u_2, u_1)~.
\end{equation}
Then, formula \eqref{eq:formula of R Matrix} implies
\begin{equation}\label{eq:formula of matrix form}
{\cal R}_{\vl, \vm} (u_1,  u_2 | q,t) =
\sum_{\vn} \mathcal{A}^{\mathrm{op}}_{\vn,\vl}(u_1, u_2)
\mathcal{A}^{-1}_{\vm, \vn}(u_1, u_2).
\end{equation}
We have calculated the $R$-matrix using \eqref{eq:formula of matrix form}
up to $|\vl| \leq 3$ and checked that the Yang-Baxter equation is satisfied.

More generally if we employ the original normalization of the
generalized Macdonald function, $k_{\lambda\mu}$ in \eqref{Rdef} are
regarded as the diagonal elements of the $R$-matrix.  In this case, we
have the following formula:
\begin{equation}\label{eq:generalization}
{\cal R}_{\vl, \vm} (u_1,  u_2 | q,t) =
\sum_{\vn} \mathcal{A}^{\mathrm{op}}_{\vn,\vl}(u_1, u_2)
{\cal R}_{\vn}^{\rm diag}
\mathcal{A}^{-1}_{\vm, \vn}(u_1, u_2),
\end{equation}
where ${\cal R}_{\vn}^{\rm diag}=k_{\vn}$ is
a diagonal $R$-matrix.
\eqref{eq:generalization} means the generalized Macdonald
polynomial diagonalizes the DIM $R$-matrix.
The formula \eqref{eq:generalization} should be compared with
the factorization of the universal $R$-matrix of the
quantum affine algebra
\eqref{eq:43} in Appendix E.

\paragraph{Appendix C: Anomaly cancellation by weaving.}
In this Appendix, we describe in detail how to modify the intertwiners
of the DIM algebra to cancel the anomaly. As we mentioned in
sec.~8, the necessary modification involves tensor products with extra
factors, which we denote by $\widetilde{\Psi}^{\varnothing}$ and
$\widetilde{\Psi}^{*}_{\varnothing}$, depending on extra scalar
fields. These factors are given by the formulas similar to
Eq.~\eqref{eq:16} with the only difference being the
factor of $i$ in the exponential:
\begin{equation}
  \label{eq:68}
  \widetilde{\Psi}_{\varnothing}(z) = :\prod_{k \geq 0}
  e^{-i\widetilde{\phi}(q^k z)}: \qquad \qquad
  \widetilde{\Psi}^{*}_{\varnothing}(w) = :\prod_{k \geq 0}
  e^{i\widetilde{\phi}^{*}(q^k w)} :
\end{equation}
Indeed, tensoring the intertwiners of the DIM algebra with auxiliary
$\Psi$-factors,
\begin{align}
  \parbox{1.5cm}{\includegraphics[width=1.5cm]{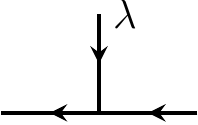}}
  = \Psi^\lambda(z) \to \Psi^\lambda(z) \otimes
  \widetilde{\Psi}^{\varnothing}(z)
  &=   \parbox{8cm}{\includegraphics[width=8cm]{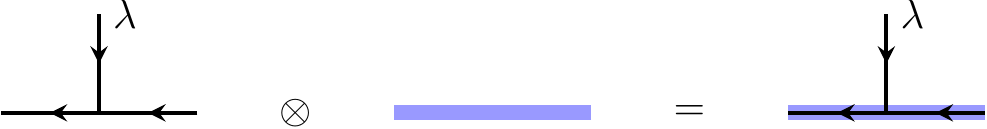}}\\
  {}\notag \\
    \parbox{1.5cm}{\includegraphics[width=1.5cm]{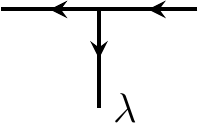}}
  = \Psi^*_\lambda(w) \to
  \Psi^*_\lambda(w) \otimes \widetilde{\Psi}^*_{\varnothing} (w)& =   \parbox{8cm}{\includegraphics[width=8cm]{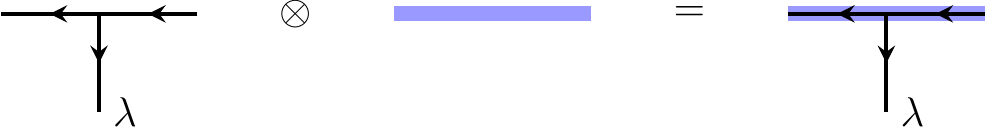}}
\end{align}
and constructing the $T$-operator via the same formula (\ref{eq:3}),
one cancels the scalar factor $E_{q,t}(z_1, w_1, z_2, w_2)$. Thus,
this new $T$-operator satisfies the \emph{standard} $RTT$ relation and
is a true group element in the Fock representation.

This modification does not eliminate the anomaly in the
\emph{perpendicular} channel, in which the additional
$\widetilde{\Psi}$-operators have no effect. To cancel this
perpendicular anomaly, one has to introduce \emph{one more} tensor
factor, spectral dual of the first one, to the
intertwiner. Eventually, we have:
\begin{gather}
   \Psi^\lambda(u|z) \to \Psi^{\lambda, \mu}_{\mathrm{tot}}(u|z) = \Psi^\lambda(u|z) \otimes
  \widetilde{\Psi}^{\varnothing}(u|z) \otimes \left(
    \widetilde{\widetilde{\Psi}}{}^{\mu}(u|z)  | \varnothing
    \rangle \right) =\\
  =   \parbox{11cm}{\includegraphics[width=11cm]{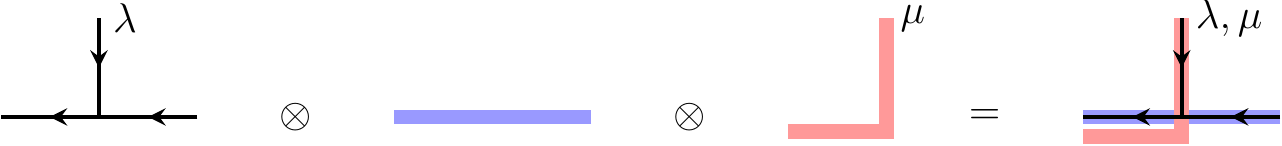}}\\
  {}\notag \\
  \Psi^*_\lambda(-uw|w) \to \Psi^{\mathrm{tot}*}_{\lambda, \mu}(-uw|w) = \Psi^*_\lambda(-uw|w) \otimes
  \widetilde{\Psi}^*_{\varnothing} (-uw|w) \otimes \left( \langle
    \varnothing | \widetilde{\widetilde{\Psi}}{}^*_{\mu} (-uw|w) \right)
  =\\
  =  \parbox{11cm}{\includegraphics[width=11cm]{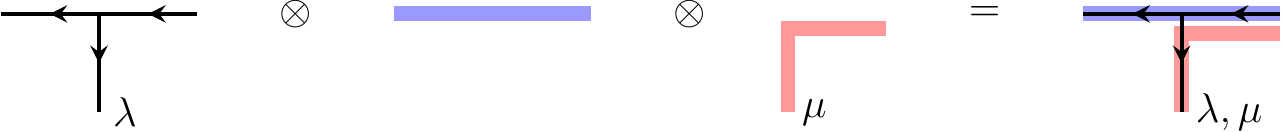}}
\end{gather}
Notice that here the DIM algebra elements act \emph{only} in the first
tensor component, as $\widetilde{\Psi}$ and
$\widetilde{\widetilde{\Psi}}$ do not satisfy the intertwining
property because of the $i$ in the exponential.

One can picture the extra tensor factors added to the network matrix
model as a woven fabric with $\widetilde{\Psi}$ forming the horizontal
threads (warp) and $\widetilde{\widetilde{\Psi}}$ forming the vertical ones
(weft). The non-anomalous $RTT$ relations in both the horizontal and
vertical channels hold due to the commutation relations \emph{on each
  thread}, while the threads themselves trivially commute, i.e.\ can
be interwoven using just the permutation operator.
\begin{figure}[h]
  \centering
\includegraphics[width=7cm]{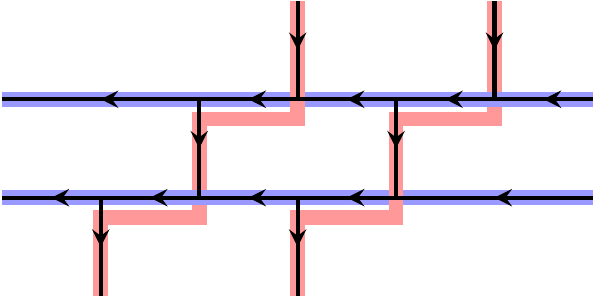}
\caption{The modified network matrix model as a weave with extra
  tensor factors shown in blue (warp threads) and red (weft threads).}
  \label{fig:3}
\end{figure}
Summarizing, the anomaly in the $RTT$ relations can be cancelled at
the expense of adding extra factors, which do not transform under the DIM
algebra, to each Fock representation.

\paragraph{Appendix D: The $4d$ limit.}
To understand the relation between network matrix models and familiar
objects in CFT, we discuss in this Appendix the ``$4d$ limit'': $q =
e^{\hbar}\to 1$, $t = q^{\beta}$, $\beta$ fixed. Algebraically, it is
described by the affine Yangian, \cite{Yangian}--\cite{Matsuo}.  In
this limit, the intertwiners should turn into the {\it screened}
vertex operators of the ordinary Virasoro or $\mathcal{W}_m$-algebra.
We demonstrate here how this happens in detail.

In the $4d$ limit, the $q$-deformation of the free field disappears
and one has\footnote{The parameter $\beta$ can be eliminated from the
  commutation relations by an overall rescaling of the field
  $\phi(z)$.}
\begin{equation}
\label{eq:25}
  \left[a_n, a_m\right] = \frac{n}{\beta} \delta_{m+n,0}
\end{equation}
The DIM currents $\eta(z)$ and $\xi(z)$ (for fixed $z$) turn into
exponentials of the ordinary Heisenberg currents:
\begin{gather}
  \label{eq:18}
  \eta(z) \to :e^{\hbar \beta z \partial_z \phi(z)}:\\
   \xi(z)\to :e^{-\hbar \beta z \partial_z \phi(z)}:
\end{gather}
To get a nontrivial result for the vertices $\Psi_{\lambda}$ and
$\Psi^{*}_{\lambda}$, one has to assume that the rows $\lambda_i$ of
the Young diagram become longer and longer in the limit of $\hbar \to
0$, so that $\hbar \lambda_i$ is finite. It is also important that the
\emph{number} of rows $l(\lambda)$ remains finite. Then, the products
over rows in Eqs.~\eqref{eq:23},~\eqref{eq:24} become exponentials of
integrals:
\begin{gather}
  \label{eq:19}
  :\prod_{(i,j)\in \lambda} \eta(z q^{j-1} t^{1-i}): \to
  :\prod_{i=1}^{l(\lambda)} \exp \left( \beta \int_1^{pe^{\hbar
        \lambda_i}} z \partial_z \phi(z w) \frac{dw}{w}\right):
  = :\prod_{i=1}^{l(\lambda)} e^{\beta \phi(x_i) - \beta \phi(z)}: \\
  :\prod_{(i,j)\in \lambda} \xi(z q^{j-1} t^{1-i}): \to
  :\prod_{i=1}^{l(\lambda)} e^{-\beta \phi(x_i) + \beta
    \phi(z)}:\label{eq:17}
\end{gather}
where $x_i = z e^{\hbar \lambda_i}$. Thus, the essential part of the
vertex is just a product of the Dotsenko-Fateev (DF) screening currents. Notice
that the number of screening currents is \emph{not fixed,} but can be
arbitrary, depending on the height of $\lambda$. In the end, when
computing the topological string partition function, or the conformal block,
one should sum over $\lambda$. In the $4d$ limit, the role of this sum
is twofold: it produces a multiple \emph{integral} over $x_i$, but
also gives a \emph{sum} over the number of the DF screening charges. This
sum is customary in the DF formalism, since one has to accommodate for
any external and internal dimensions in the conformal block.

The part of the vertex independent of the Young diagram
$\Psi_{\varnothing}(z)$ diverges in the $4d$ limit. More precisely,
one gets:
\begin{align}
  \label{eq:7}
  \Psi_{\varnothing}(z) &\to \exp \left( \frac{1}{\hbar} \int_0^z
    \phi(w) \frac{dw}{w} + \frac{1}{2}\phi(z)+ \mathcal{O}(\hbar)
  \right),\\
  \Psi_{\varnothing}^{*}(z) &\to \exp \left( - \frac{1}{\hbar} \int_0^z
    \phi(w) \frac{dw}{w} - \left( \beta - \frac{1}{2} \right)
    \phi_{-}(z) -\left( \frac{3}{2} - \beta \right) \phi_{+}(z)  + \mathcal{O}(\hbar) \right),\label{eq:22}
\end{align}
The origin of this divergence can be traced back to our assumption
that the position of the vertex $z$ remains finite. It turns out that
one has to modify this assumption to get a meaningful operator in the
$4d$ limit. The only way to cancel the divergence is to combine
$\Psi(z)$ and $\Psi^{*}(w)$ pairwise on each leg, and send the
spectral parameters inside the pairs towards each other,
e.g. $\frac{z}{w} = q^{\alpha}$ with $\alpha$ finite. Then, the
divergent parts in Eqs.~\eqref{eq:7}~\eqref{eq:22} cancel each other
and we are left with the following vertex operator:
\begin{equation}
  \label{eq:26}
  \Psi_{\varnothing}(z) \Psi_{\varnothing}^{*}(z q^{\alpha}) \to \exp
  \left[ (1-\beta - \alpha) \phi_{-}(z) + (\beta -1 -\alpha)
    \phi_{+}(z) \right]
\end{equation}
Notice that positive and negative modes enter with slightly different
coefficients in the vertex operator~\eqref{eq:26}. This leads to the
special form of the so-called $U(1)$ vertex
operators~\cite{CarlssonOkounkov}. The pairwise arrangement of the
intertwiners implies that the whole network is \emph{balanced,} i.e.\
it is composed of the blocks of intertwiners, which conserve the
slopes of the lines. The simplest block of this form is the four-leg
block, which will play the role of the $T$-operator in the
main text.

Let us also mention that for a special choice of the spectral
parameters (\emph{a la} DF, which we will use henceforth) the width of
the Young diagram $\lambda$ is limited. To see this consider the
product of two intertwiners:
\begin{equation}
  \label{eq:27}
  \parbox{3.5cm}{\includegraphics[width=3.5cm]{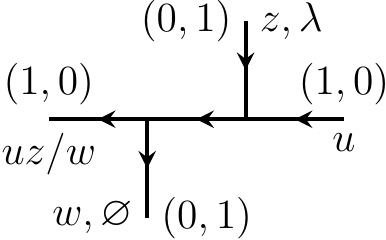}} = \Psi^{\lambda}(u|z) \Psi^{*}_{\varnothing}(-uz|w)
\end{equation}
featuring as the main element of the conformal blocks. For concrete
calculations the two operators should be normal ordered, which gives
rise to an OPE coefficient. This coefficient contains a
multiplicative factor
\begin{equation}
  \label{eq:28}
  \prod_{(i,j)\in \lambda} \left( 1 - \sqrt{\frac{q}{t}} \frac{w}{z}
    q^{j-1} t^{1-i} \right)
\end{equation}
If we set $\sqrt{\frac{q}{t}} \frac{w}{z} = t^N$ with positive integer
$N$, the contribution of Young diagrams wider than $N$ is exactly zero
because of the factor with $j=1$ and $i=N+1$ in the product. For
example we can set $N=1$ and observe that there can only be either one
or no screening currents around a given vertex operator. This is the
reflection of the situation in the original DF setup for conformal
block, where the possible number of screenings is governed by the
dimensions of the (degenerate) external fields. In Eq.~\eqref{eq:27}
we have set one of vertical Young diagrams empty. If we consider
general pairs of diagrams $\lambda$ and $\mu$ the situation is
analogous and there is a constraint on their \emph{relative} widths.

Let us summarize what we obtained in this Appendix. In the $4d$
limit one is forced to consider asymptotically large partitions
sitting on the vertical legs. This introduces asymmetry in the
originally symmetric description of the network of intertwiners and
breaks the spectral duality symmetry. The sums over vertical Young
diagrams turn into multiple integrals, while horizontal lines still
carry free field (Fock) representations. The original DIM intertwiners
make no sense, unless considered in \emph{balanced} pairs with
coalescing spectral parameters. The \emph{pair} of vertices turns into
a prototype of the vertex operator with the product of screening
currents as follows (we omit some normalization factors in front of
the operators):
\begin{equation}
  \label{eq:29}
    \parbox{3.5cm}{\includegraphics[width=3.5cm]{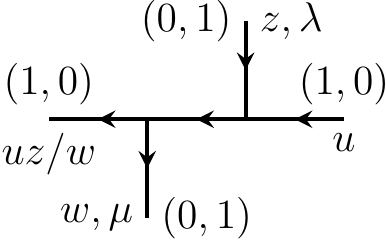}} \xrightarrow{q \to 1, t =
    q^{\beta}, w = z q^{-\alpha}}   :e^{
   (1-\beta - \alpha) \phi_{-}(z) + (\beta -1 -\alpha)
    \phi_{+}(z) } \prod_{i=1}^{l(\mu)} e^{-\beta \phi(x_i) + \beta
    \phi(z)} \prod_{i=1}^{l(\lambda)} e^{\beta \phi(x_i) - \beta
    \phi(z)}:
\end{equation}

\paragraph{Appendix E: Horizontal and vertical $R$-matrices.} In this
Appendix, we give a simplified and hopefully elucidating example, in
which horizontal and vertical $R$-matrices occur.

The horizontal and vertical $R$-matrices correspond to directions in
the root space of the algebra, which parameterize the choice of the
Borel subalgebra. Thus, the change from one $R$-matrix to another is
associated with different choices of the Borel subalgebra.

Consider first the finite dimensional quantum algebra
$U_q(\mathfrak{sl}_n)$. The universal $R$-matrix is given by
the product over positive roots (and the Cartan subalgebra):
\begin{equation}
  \label{eq:40}
  \mathcal{R} = e^{\frac{1}{2}\sum_{i,j} \alpha^{-1}_{ij} \alpha_{ii}
    \alpha_{jj} H_i \otimes H_j} \prod_{\alpha \in \Delta_{+}}
  \exp_{q^{(\alpha,\alpha)}} \left( (q-q^{-1}) E_{\alpha} \otimes E_{-\alpha} \right)
\end{equation}
where $\alpha_{ij}$ is the Cartan matrix of $A_{n-1}$ and the
$q$-exponential is given by
\begin{equation}
  \label{eq:42}
  \exp_q (x) = \sum_{n \geq 0} \frac{(1-q)^n}{\prod_{k=1}^n(1-q^k)} x^n
\end{equation}
This $R$-matrix belongs to the tensor product of positive and negative
Borel subalgebras $U_q(\mathfrak{b}_{+}) \otimes
U_q(\mathfrak{b}_{-})$ and in this sense is a triangular matrix. In
the fundamental representation, one gets the matrix (up to an overall
constant):
\begin{equation}
  \label{eq:38}
  \mathcal{R}_{\square, \square} = (\rho_{\square} \otimes \rho_{\square}) \mathcal{R} =  q \sum_{i=1}^n  E_{ii} \otimes E_{ii} + \sum_{i \neq j} E_{ii}
  \otimes E_{jj} + (q-q^{-1}) \sum_{i<j} E_{ij} \otimes E_{ji}
\end{equation}
Notice the $i<j$ constraint in the last sum, which defines
$\mathfrak{b}_{\pm}$.

If we change the Borel subalgebras $\mathfrak{b}_{\pm}$ that enter
the definition of the universal $R$-matrix, i.e.\ rotate the
hyperplane separating positive and negative roots, we get a different
$R$-matrix:
\begin{equation}
  \label{eq:41}
  \mathcal{R}^{\text{rot}} =   e^{\frac{1}{2}\sum_{i,j} \alpha^{-1}_{ij} \alpha_{ii}
    \alpha_{jj} H_i \otimes H_j} \prod_{\alpha \in \Delta^{\text{rot}}_{+}}
  \exp_{q^{(\alpha,\alpha)}} \left( (q-q^{-1}) E_{\alpha} \otimes E_{-\alpha} \right)
\end{equation}
However, this $R$-matrix is obtained from $\mathcal{R}$ by a
simple transformation. One should conjugate $\mathcal{R}$ with the
element of the Weyl group, which performs the rotation of the
hyperplane in root space:
\begin{equation}
  \label{eq:36}
  \mathcal{R}^{\text{rot}} = \mathcal{R}^{(\sigma)} = (\sigma \otimes \sigma) \mathcal{R} (\sigma^{-1} \otimes \sigma^{-1})
\end{equation}
In the fundamental representation, the rotated $R$-matrix reads
\begin{equation}
  \label{eq:39}
  \mathcal{R}^{\text{rot}}_{\square, \square} = (\rho_{\square} \otimes \rho_{\square}) \mathcal{R}^{\text{rot}} =  q \sum_{i=1}^n  E_{ii} \otimes E_{ii} + \sum_{i \neq j} E_{ii}
  \otimes E_{jj} + (q-q^{-1}) \sum_{i<j} E_{\sigma(i)\sigma(j)} \otimes E_{\sigma(j)\sigma(i)}
\end{equation}
Obviously $\mathcal{R}^{\text{rot}}$ is equivalent to $\mathcal{R}$ in
all respects, in particular, if $\mathcal{R}$ satisfies the
Yang-Baxter equation, so does $\mathcal{R}^{\text{rot}}$.

\begin{figure}[h]
  \centering
  \includegraphics[width=8cm]{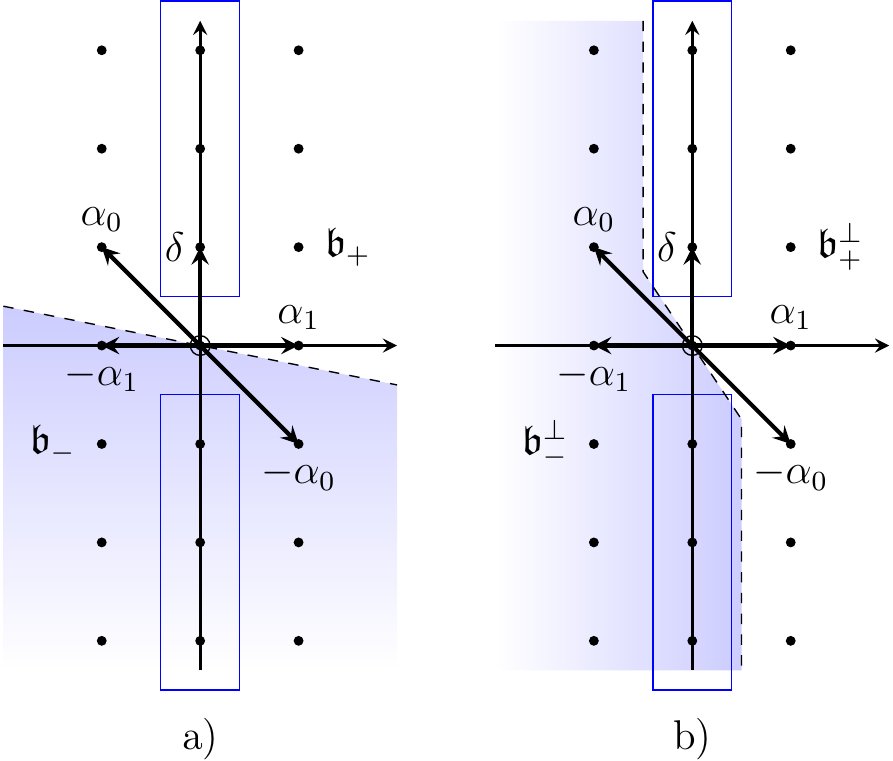}
  \caption{a) The root system of $U_q(\widehat{\mathfrak{sl}}_2)$ with
    the standard choice of positive roots $\alpha_1$ and $\alpha_0 =
    \delta - \alpha_1$ and the corresponding ``vertical'' Borel
    subalgebras $\mathfrak{b}_{-}$ (shaded) and $\mathfrak{b}_{+}$. b)
    The ``horizontal'' Borel subalgebras
    $\mathfrak{b}^{\perp}_{\pm}$. The modes of the Cartan current
    $H(z)$ are shown in blue boxes.}
  \label{fig:2}
\end{figure}

In the case of quantum \emph{affine} algebras, the situation is a bit
more subtle. The $R$-matrix is again given by the product over
positive roots belonging to the Borel subalgebra $\mathfrak{b}_{+}$
shown in Fig.~\ref{fig:2}, a), though now the product is infinite:
\begin{equation}
  \label{eq:43}
  \mathcal{R} = \mathcal{K} \mathcal{R}_{>\delta} \mathcal{R}_{\sim
    \delta} \mathcal{R}_{< \delta}
\end{equation}
where
\begin{gather}
  \label{eq:44}
  \mathcal{K} = e^{\frac{1}{2}\sum_{i,j} \alpha^{-1}_{ij} \alpha_{ii}
    \alpha_{jj} H_i \otimes H_j}\\
  \mathcal{R}_{>\delta} = \prod_{\alpha \in \Delta_{+}} \prod_{n \geq 0}
  \exp_{q^{(\alpha,\alpha)}} \left( (q-q^{-1}) E_{\alpha+n\delta}
    \otimes E_{-\alpha-n\delta} \right)\\
  \mathcal{R}_{\sim \delta} = \exp \left( (q-q^{-1}) \sum_{n \geq 1} \sum_{i,j} u_{n,i,j}
    E_{n\delta,i}
    \otimes E_{-n\delta,j} \right)\\
  \mathcal{R}_{<\delta} = \prod_{\alpha \in \Delta_{-}} \prod_{n \geq 1}
  \exp_{q^{(\alpha,\alpha)}} \left( (q-q^{-1}) E_{\alpha+n\delta}
    \otimes E_{-\alpha-n\delta} \right)
\end{gather}
and $u_n$ is inverse of the matrix $(-1)^{n(1-\delta_{ij})} [n
a_{ij}]_q n^{-1}$. In the fundamental evaluation representation of
$U_q(\widehat{\mathfrak{sl}}_2)$, the parts of $R$-matrix up to a
scalar multiple look as follows:
\begin{gather}
  \label{eq:45}
  \rho_{z,\square}\otimes \rho_{w,\square} (\mathcal{K}) = \left(
    \begin{array}{cccc}
      1 & 0 & 0 & 0\\
      0 & q^{-1} & 0 & 0\\
      0 & 0 & q^{-1} & 0\\
      0 & 0 & 0 & 1\\
    \end{array}
\right), 
  \qquad \qquad
  \rho_{z,\square}\otimes \rho_{w,\square} (\mathcal{R}_{>\delta}) = \left(
    \begin{array}{cccc}
      1 & 0 & 0 & 0\\
      0 & 1 & \frac{q-q^{-1}}{1-\frac{z}{w}} & 0\\
      0 & 0 & 1 & 0\\
      0 & 0 & 0 & 1\\
    \end{array}
\right) 
  \\
  \rho_{z,\square}\otimes \rho_{w,\square} (\mathcal{R}_{\sim\delta})
  = \left(
    \begin{array}{cccc}
      1 & 0 & 0 & 0\\
      0 & \frac{1 - q^2 \frac{z}{w}}{1 - \frac{z}{w}} & 0 & 0\\
      0 & 0 & \frac{1-\frac{z}{w}}{1 - \frac{z}{q^2w}} & 0\\
      0 & 0 & 0 & 1\\
    \end{array}
\right),  
  \qquad
  \rho_{z,\square}\otimes \rho_{w,\square} (\mathcal{R}_{<\delta}) = \left(
    \begin{array}{cccc}
      1 & 0 & 0 & 0\\
      0 & 1 & 0 & 0\\
      0 & \frac{(q-q^{-1})\frac{z}{w}}{1-\frac{z}{w}} & 1 & 0\\
      0 & 0 & 0 & 1\\
    \end{array}
\right) 
\end{gather}
Thus, the whole $R$-matrix is of the familiar form:
\begin{equation}
  \label{eq:46}
R \left( \frac{z}{w} \right) = \rho_{z,\square}\otimes \rho_{w,\square} (\mathcal{R}) = \left(
    \begin{array}{cccc}
      1 & 0 & 0 & 0\\
      0 & \frac{w - z}{ q w - q^{-1} z} &
      \frac{(q - q^{-1})w}{q w - q^{-1} z} & 0\\
      0 & \frac{(q-q^{-1})z}{q w-q^{-1} z} &
      \frac{w - z}{q w -q^{-1} z} & 0\\
      0 & 0 & 0 & 1\\
    \end{array}
\right)
\end{equation}
Of course, the form of the $R$-matrix again depends on the choice of
the Borel subalgebra $\mathfrak{b}_{+}$. First of all, there is an
infinite set of $R$-matrices $\mathcal{R}^{\mathrm{rot}}$ obtained
from a given one by action of the affine Weyl group. They are given by
expressions similar to Eq.~\eqref{eq:36}.

Most importantly, there is one distinct choice of the Borel subalgebra
denoted by $\mathfrak{b}^{\perp}$ which cannot be obtained from the
standard one by the action of the Weyl group: one can choose the
subalgebra ``in the perpendicular direction'' (see Fig.~\ref{fig:2},
b)). This choice is natural in Drinfeld's ``new realization'' of
quantum affine algebras and can be thought of as a limiting case, when
one acts with the $T_{\theta}^n$ generator of the affine Weyl group
with a sufficiently high power $n$. Let us compute this
``perpendicular'' $R$-matrix in the evaluation representation. What we
need is just a minor variation of the Khoroshkin-Tolstoy
expression~\eqref{eq:43}. The $R$-matrix is given by the product
\begin{equation}
  \label{eq:37}
  \mathcal{R}^{\perp} = \mathcal{K} \mathcal{R}_{> \delta}' \mathcal{R}_{\sim \delta}
\end{equation}
Notice that there is no usual $R_{<\delta}$ term in the product and
instead, the $R_{>\delta}$ term is modified:
\begin{equation}
  \label{eq:48}
  \mathcal{R}'_{>\delta} = \prod_{\alpha \in \Delta_{+}} \prod_{n \in \mathbb{Z}}
  \exp_{q^{(\alpha,\alpha)}} \left( (q-q^{-1}) E_{\alpha+n\delta}
    \otimes E_{-\alpha-n\delta} \right)
\end{equation}
Performing the calculation, we find
\begin{gather}
  \label{eq:47}
    \rho_{z,\square}\otimes \rho_{w,\square} (\mathcal{R}_{>\delta}') = \left(
    \begin{array}{cccc}
      1 & 0 & 0 & 0\\
      0 & 1 & (q-q^{-1})\delta \left( \frac{z}{w} \right) & 0\\
      0 & 0 & 1 & 0\\
      0 & 0 & 0 & 1\\
    \end{array}
\right)
\end{gather}
where $\delta(x) = \sum_{n \in \mathbb{Z}} x^n$. Multiplying this
modified matrix with $\mathcal{R}_{\sim \delta}$, one finds that the
$\delta$-function term is annihilated since $(1-x) \delta(x) = 0$ and the
perpendicular $R$-matrix is diagonal:
\begin{equation}
  \label{eq:49}
  R^{\perp}\left( \frac{z}{w} \right)  = \rho_{z,\square}\otimes \rho_{w,\square}(\mathcal{R}^{\perp}) =
  \rho_{z,\square}\otimes \rho_{w,\square}(\mathcal{K} \mathcal{R}_{\sim \delta})
  = \left(
    \begin{array}{cccc}
      1 & 0 & 0 & 0\\
      0 & \frac{q^{-1} w - q z}{w - z} & 0 & 0\\
      0 & 0 & \frac{w-z}{q w - q^{-1}z} & 0\\
      0 & 0 & 0 & 1\\
    \end{array}
\right)
\end{equation}
$\mathcal{R}^{\perp}$ trivially satisfies the Yang-Baxter equation and
the relation $\mathcal{R^{\perp}} (x^{-1}) = P(\mathcal{R}^{\perp} (x))^{-1}P
$. The $R$-matrix~\eqref{eq:49} differs from the familiar
one~\eqref{eq:46} by a twist, which is nothing, but the product over a
``quarter'' of the roots~\cite{Drinnew,EKP}.

There are several lessons to learn from this example. $\mathcal{R}$
and $\mathcal{R}^{\perp}$ are very similar to the $R$-matrices of the
DIM algebra taken in the basis of Macdonald and generalized Macdonald
polynomials respectively. The natural basis of weight vectors $\left|
  z, \pm \frac{1}{2} \right\rangle $ in the evaluation representation
is a counterpart of the generalized Macdonald basis. Indeed,
generalized Macdonald polynomials are eigenvalues of
$\Delta(x^{+}_0)$, which can be thought of as a Cartan generator of
DIM. Similarly, we have
\begin{equation}
  \label{eq:50}
  \rho_{z,\square}\otimes \rho_{w,\square} (\Delta^{D} (H_1) ) \left| z, \pm \frac{1}{2} \right\rangle \otimes
  \left| w, \pm \frac{1}{2} \right\rangle = (\pm z \pm w) \left| z, \pm
    \frac{1}{2} \right\rangle \otimes
  \left| w, \pm \frac{1}{2} \right\rangle
\end{equation}
where $H_1$ is the first mode of the Drinfeld current $H(z)$, which is
the Cartan generator. Notice that here $\Delta^D$ is not the standard
coproduct $\Delta$, which would have led to the standard
$R$-matrix~\eqref{eq:46}, but the \emph{second}
coproduct~\cite{Drinnew,FJMMl}, which is also known as the Drinfeld
coproduct.

The evaluation representation can be thus interpreted as ``vertical'',
since we can diagonalize the ``vertical'' Cartan generators $H(z)$ in
it. What are the ``horizontal'' representations? It seems natural that
those are the highest weight representations of the affine
algebra. It is technically difficult to derive the $R$-matrix in these
representations. However, we hope that if computed, this $R$-matrix,
among other things, can be used to obtain new and interesting knot
invariants.

\section*{Acknowledgements}

We are grateful to Michio Jimbo for enlightening advices and to Yutaka Matsuo for numerous discussions.

Our work is supported in part by Grant-in-Aid for Scientific Research
(\# 24540210) (H.A.), (\# 15H05738) (H.K.), for JSPS Fellow (\#
26-10187) (Y.O.) and JSPS Bilateral Joint Projects (JSPS-RFBR
collaboration) ``Exploration of Quantum Geometry via Symmetry and
Duality'' from MEXT, Japan. It is also partly supported by grants
15-31-20832-Mol-a-ved (A.Mor.), 15-31-20484-Mol-a-ved (Y.Z.),
16-32-60047-Mol-a-dk (And.Mor), by RFBR grants 16-01-00291 (A.Mir.),
15-01-09242 (An.Mor.) and 14-01-00547 (Y.Z.), by joint grants
15-51-50034-YaF, 15-51-52031-NSC-a, 16-51-53034-GFEN,
16-51-45029-IND-a. The work of Y.Z. was supported in part by INFN and
by the ERC Starting Grant 637844-HBQFTNCER.

\end{document}